\documentclass[preprint,journal]{vgtc}            

\onlineid{2032}

\vgtccategory{Representations \& Interaction}

\title{Super-Gaussian: Interactive Scene Editing for 3D Gaussian Splatting and NLI-Based Volume Visualization in Virtual Reality}
\author{%
  \authororcid{Suemin Jeon}{0009-0007-1013-4845},
  \authororcid{Kaiyuan Tang}{0009-0001-3512-0112},
  \authororcid{Chaoli Wang}{0000-0002-0859-3619},
  and
  \authororcid{Won-Ki Jeong}{0000-0002-9393-6451}
}

\authorfooter{
  \item
  	Suemin Jeon and Won-Ki Jeong are with Korea University.\\
  	E-mail: \{orangeblush, wkjeong\}@korea.ac.kr
  \item
  	Kaiyuan Tang and Chaoli Wang are with the University of Notre Dame.\\
  	E-mail: \{ktang2, chaoli.wang\}@nd.edu
}
\teaser{
  \centering
    \includegraphics[width=\linewidth]{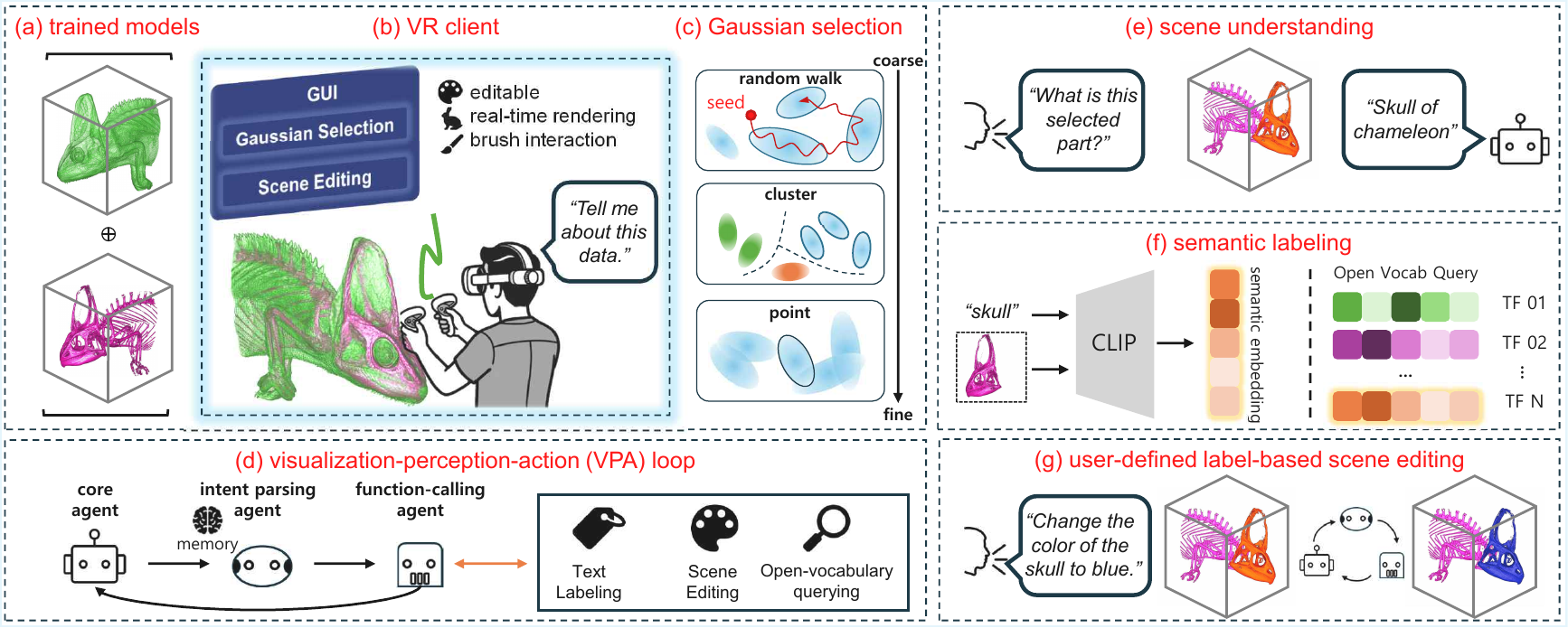}
  \vspace{-0.3in}
  \caption{The pipeline of our Super-Gaussian framework. (a) \sm{One or more }editable 3DGS models trained from multi-view images. (b) A VR client that enables users to explore volumetric data and perform real-time editing and selection through brush interaction. (c) Given brush input, the system supports three selection methods: random walk, cluster selection, and point selection, allowing flexible ROI specification. (d) The selected parts can be further interpreted or edited via NLI through a hierarchical agent comprising a core agent, an intent-parsing agent, and a function-calling agent, thereby enabling the VPA loop. (e)  Selected regions can be queried for scene understanding via NLI, allowing users to interpret volumetric structures through open-vocabulary questions. (f) User-selected regions can be assigned semantic text labels on the fly, enabling open-vocabulary querying via CLIP-based embeddings. (g) User-defined labels enable semantic scene editing via natural-language commands.
  \label{fig:teaser}
  \vspace{-0.05in}
    }
}
\abstract{
Despite the promise of virtual reality (VR) for intuitive spatial interaction, volume visualization (VolVis) in VR remains constrained by high rendering costs and motion discomfort. 
Recent advances have shown that representing volumetric scenes with 3D Gaussian splatting enables high-performance rendering, mitigating the computational latency typically associated with immersive 3D data exploration, thus making this representation well-suited for VR. 
However, existing Gaussian-based scene editing workflows remain limited by slow offline segmentation and fatigue-inducing manual selection. 
To address these challenges, we present Super-Gaussian, a novel VolVis framework that enhances scene editing and interaction in VR through intuitive 3D Gaussian selection and natural language interaction (NLI). 
Our approach groups Gaussian primitives into higher-level units via feature-aware clustering, enabling efficient selection of complex volumetric regions, such as tumors in medical images or filaments in cosmological data, without tedious point-by-point interaction. 
Building on this representation, we introduce a hierarchical select-and-refine workflow that combines random-walk-based region propagation, cluster selection, and point refinement, allowing users to progressively specify regions of interest with reduced effort. 
We further support on-the-fly text labeling of user-selected regions using NLI, allowing users to semantically query, interpret, and manipulate selected content within a visualization–perception–action loop. 
By integrating multimodal interaction, including speech, visual feedback, and spatial manipulation in VR, our framework supports intuitive and flexible exploration, editing, and scientific analysis of volumetric data. 
We demonstrate the effectiveness of Super-Gaussian through four representative case studies, \sm{quantitative selection benchmarks against existing Gaussian-based selection techniques, and system-level evaluations.}
\sm{Implementation details, experiments, and source code can be found on the project page:
\url{https://smin0136.github.io/super-gaussian-project/}.}
}


\graphicspath{{figs/}{figures/}{pictures/}{images/}{./}} 

\usepackage{tabu}                      
\usepackage{booktabs}                  
\usepackage{lipsum}                    
\usepackage{mwe}                       
\usepackage{ccicons}                   
\usepackage{amsmath}
\usepackage{bm}
\usepackage{makecell}

\usepackage{amsfonts}
\usepackage{adjustbox}
\usepackage{comment}
\usepackage{multirow}

\newenvironment{myitemize}{
\begin{itemize}
 \setlength{\itemsep}{1pt}
 \setlength{\parskip}{0pt}
 \setlength{\parsep}{0pt}}{\end{itemize}
}

\usepackage{xcolor}
\newcommand{\sm}[1]{{\color{black}#1}}

\begin{document}



\firstsection{Introduction}

\maketitle

Scientific visualization (SciVis) enables domain experts to explore, understand, and analyze complex volumetric data by revealing spatial structures within 3D fields. Volume visualization (VolVis) plays a central role in this process by allowing users to inspect internal structures and relationships within volumetric datasets. Traditionally, most VolVis systems have been designed for 2D desktop environments, where users interact with volumetric scenes using a mouse and keyboard. However, interacting with complex 3D volumetric structures through 2D interfaces can be cumbersome. In particular, layered and semi-transparent volumetric structures are often difficult to interpret through 2D interaction, limiting depth perception and slowing spatial exploration, thereby making them unintuitive.

To address these limitations, several studies have explored VolVis in virtual reality (VR)~\cite{goswami2025holoview, taibo2024immersive, boorboor2023voxar, shetty2011immersive, pidhorskyi2018syglass, yuan2023meinvr, jadhav2022md, xu2024realtime_dvr}, which offers natural spatial interaction, accurate depth perception, and intuitive exploration of 3D structures. In immersive environments, users can directly manipulate volumes through hand gestures and spatial interaction, enabling a more natural understanding of volumetric structures. However, rendering volumetric datasets directly in VR remains computationally expensive. Traditional direct volume rendering (DVR)-based rendering pipelines often suffer from low frame rates in immersive environments, limiting interactive scene editing and exploration~\cite{goswami2025holoview, taibo2024immersive, boorboor2023voxar, shetty2011immersive, pidhorskyi2018syglass,  xu2024realtime_dvr, meng2020eye,waschk2020favr,usher2017virtual}. This issue is particularly problematic in VR, where frame-rate drops can cause motion discomfort and significantly degrade user experience~\cite{laviola2000discussion, scholl2018direct}. As a result, despite the advantages of immersive interaction, exploring large volumetric datasets in VR has remained a significant challenge.

Recent advances in 3D Gaussian splatting (3DGS)~\cite{kerbl2023gs} have introduced a new paradigm for representing volumetric scenes. By modeling scenes as collections of explicit Gaussian primitives, 3DGS enables compact scene representations and highly efficient rendering that is largely independent of the original volume resolution. These properties have led to increasing interest in Gaussian-based approaches for VolVis. 
In particular, recent work has explored the integration of editable Gaussian representations into interactive visualization pipelines. 
For instance, systems such as iVR-GS~\cite{tang2025ivrgs} and related methods 
extend 3DGS to support interactive scene editing~\cite{tang2025texgs}, allowing users to modify visualization attributes such as color, opacity, and lighting. 
Furthermore, combining Gaussian representations with natural language interaction (NLI) and large language models (LLMs)~\cite{ai2025nli4volvis} enables more intuitive scene understanding and manipulation through open-vocabulary queries and semantic editing.

Although existing editable 3DGS-based volume representations offer a promising basis for fast, efficient VolVis in immersive VR, several challenges remain. 
Many systems rely on models trained on rendered images via semantic scene segmentation, which limits how users can interact with a scene. 
Because scene editing can be performed at the level of individual Gaussian scenes, fine-grained scene editing requires retraining a new Gaussian scene from scratch with modified semantic scene segmentation. 
Although several methods have been proposed for segmenting Gaussian primitives~\cite{chen2024gaussianeditor, jiang2024vr,liu2024infusion, mirzaei2024reffusion, zhuang2024tip}, they typically depend on offline processing that can take several minutes and are generally limited to object-level segmentation, making them ill-suited for interactive exploration in VR.


Motivated by these advances and challenges, we explore how editable Gaussian-based volume representation can be extended to immersive VR for interactive scene editing and exploration. 
%
%
Our system supports both GUI-based interaction and speech-driven NLI, allowing users to issue open-vocabulary commands to interact with volumetric scenes via spatial gestures such as scaling, rotation, and navigation.
Unlike methods that rely on pre-segmentation pipelines~\cite{ai2025nli4volvis}, our framework allows users to define regions of interest (ROIs) on the fly via spatial input in VR. 
Prior Gaussian selection approaches~\cite{supersplat2024, shen2025gaussianshopvr} typically rely on point-wise primitive selection or 2D desktop interaction, which can be imprecise, repetitive, and fatiguing. 
To address this issue, we introduce \textbf{Super-Gaussian}, a feature-aware clustering of Gaussian primitives that serves as a higher-level interaction unit for volumetric editing. This representation enables semi-automatic selection of meaningful volumetric regions without predefined semantic categories and supports an iterative select-and-refine workflow for accurate, intention-driven ROI specification.
%
By integrating immersive VR interaction, NLI, and Super-Gaussian-based selection, \sm{our framework extends existing editable Gaussian-based VolVis
toward more intuitive and flexible exploration and editing of
volumetric scenes.}
Our contributions are: 

\begin{myitemize}
\vspace{-0.05in}

\item \textbf{VR-based NLI VolVis with \sm{editable} 3DGS}:\
\sm{ We present a VR framework for interactive volumetric scene editing through editable 3DGS.}
\sm{Building on existing editable 3DGS and NLI-based VolVis approaches, we couple a speech-based multi-agent NLI with VR spatial interaction, enabling intuitive spatial exploration, real-time scene editing, and natural language-driven VolVis.}

\item \textbf{Real-time ROI interaction without pre-segmentation}:\
Unlike existing Gaussian-based VolVis methods that rely on pre-segmented components or offline segmentations, our system allows users to directly specify ROIs in real time through 3D spatial interaction in VR. These user-defined regions can be annotated, edited, and incorporated into a visualization–perception–action (VPA) loop for interactive exploration.

\item \textbf{Super-Gaussian for structure-aware selection}:\
We introduce Super-Gaussian, a feature-based clustering of Gaussian splats that enables semi-automatic ROI selection. Grouping Gaussians into structure-aware units reduces repetitive, inaccurate manual selection and supports a hierarchical select-and-refine workflow that progressively aligns selections with user intent.

\vspace{-0.05in}
\end{myitemize}

\vspace{-0.075in}
\section{Related Work}

{\bf VolVis in VR.}
Prior immersive VolVis research shows that VR improves spatial understanding of volumetric data through stereoscopic depth and direct manipulation~\cite{goswami2025holoview, taibo2024immersive, boorboor2023voxar, shetty2011immersive, pidhorskyi2018syglass, yuan2023meinvr, jadhav2022md, xu2024realtime_dvr}. Many works focus on DVR interaction, especially transfer function (TF) manipulation, including real-time TF editing~\cite{choi2024real}, intensity-windowing~\cite{taibo2024immersive}, and WYSIWYG systems~\cite{wang2020immersive}. Some also combine hand and voice interaction, though often limited to specific environments~\cite{yuan2023meinvr, jadhav2022md, liu2025coordinated}.
VR introduces challenges as volumetric data can fill the display, requiring continuous per-pixel rendering, while maintaining $\geq$90 frames per second (FPS) to avoid cybersickness~\cite{laviola2000discussion, scholl2018direct}. This constrains DVR-based methods, especially for large datasets. Acceleration techniques~\cite{xu2024realtime_dvr} and foveated rendering~\cite{meng2020eye,waschk2020favr,usher2017virtual} help, but real-time performance remains difficult.
Hardware heterogeneity further complicates this, leading to platform-specific systems (e.g., for high-end devices)~\cite{hrycak2024investigating, hrycak2025investigation}. To improve scalability, compact and efficient representations are needed. Recent work uses Gaussian splatting for efficient VR rendering~\cite{kleinbeck2025multi}, but lacks editing support. In contrast, our work enables fast, editable Gaussian-splatting-based volume rendering with real-time interaction.


{\bf 3DGS in VolVis.}
3DGS~\cite{kerbl2023gs} has become a powerful representation for real-time rendering, with growing use in VolVis. Prior work demonstrates its efficiency, e.g., by compressing cinematic renderings for real-time visualization~\cite{niedermayr2024cinematic}, but standard 3DGS limits interactivity due to its baked-in appearance. Extensions address this by enabling editable shading~\cite{tang2025ivrgs}, decoupled geometry and appearance~\cite{tang2025texgs}, NLI~\cite{ai2025nli4volvis}, dynamic scene segmentation~\cite{yao2025volseggs}, TF-agnostic rendering~\cite{dyken2025volume}, and scalable or accelerated training~\cite{Bauer2025GSCache, han2025distributed, li2026variable, tang2026econgs}. 
Despite advances in rendering quality, editing, and scalability, these approaches largely target 2D desktop settings.
\sm{In contrast, our work extends existing editable Gaussian-based VolVis approaches to VR and combines them with structure-aware ROI selection through our Super-Gaussian representation. This enables users to define ROIs in real time, annotate and edit these user-defined regions, and incorporate them into a VPA loop for flexible NLI-driven exploration and real-time VolVis editing.}


\begin{figure}[t]
\centering
\includegraphics[width=\linewidth]{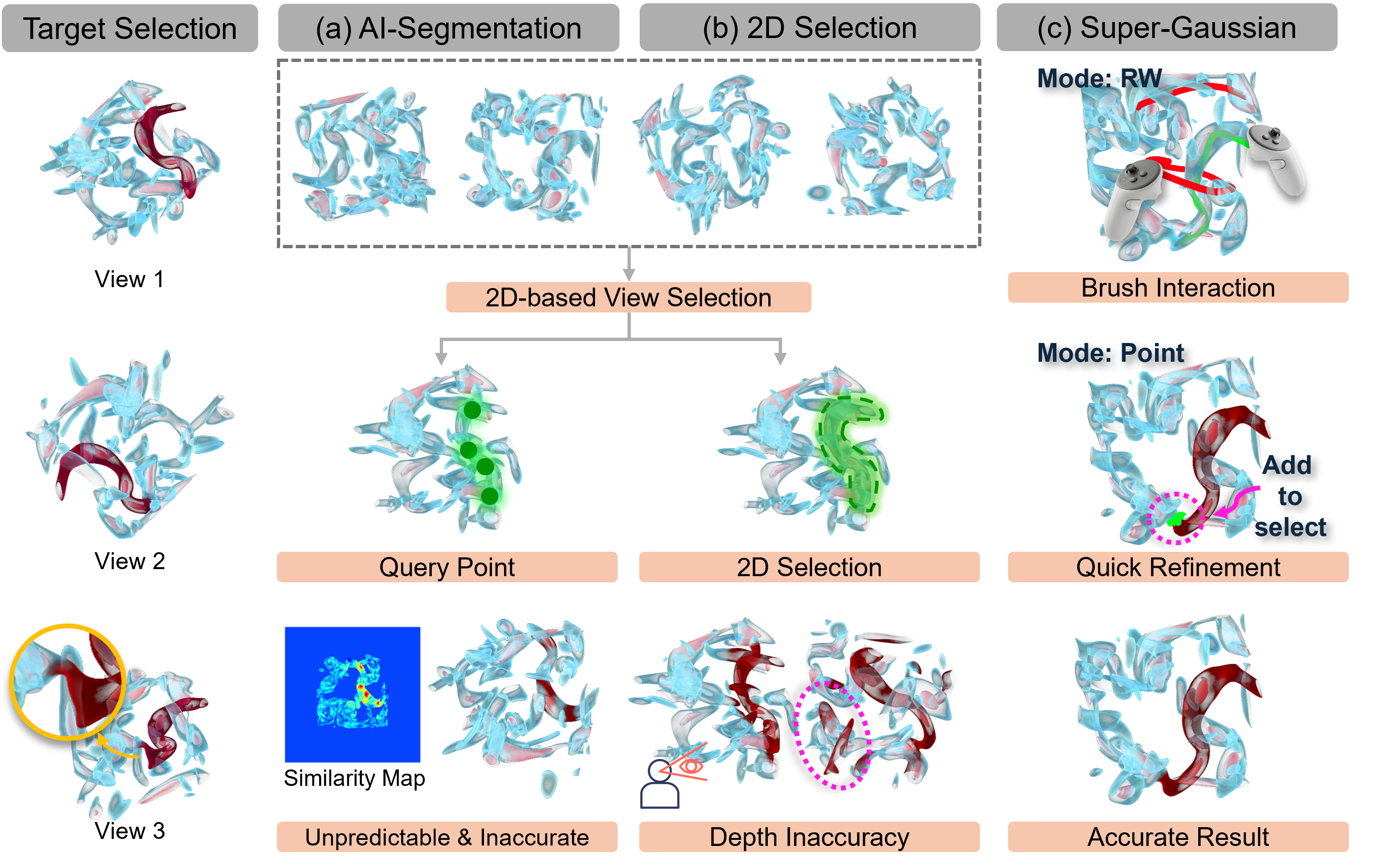}
\vspace{-0.25in}
\caption{
Comparison of Gaussian selection in semi-transparent volumetric scenes. (a) AI-based segmentation (e.g., SAGA~\cite{cen2025segment}) relies on 2D mask lifting, which requires careful view selection and often yields inconsistent 3D results. (b) Desktop-based selection (e.g., SuperSplat~\cite{supersplat2024}) uses screen-space ray casting, which suffers from depth ambiguity and imprecise selection in overlapping structures. (c) Super-Gaussian enables direct 3D brush interaction with a hierarchical select-and-refine workflow, allowing efficient, structure-aware selection with improved accuracy and reduced user effort.}
\label{fig:selection}
\vspace{-0.1in}
\end{figure}

{\bf 3D segmentation and selection.}
Because 3DGS represents scenes as independently editable primitives, tasks like inpainting~\cite{liu2024infusion,mirzaei2024reffusion}, deletion~\cite{wang2024learning}, stylization~\cite{liu2024stylegaussian}, and semantic manipulation~\cite{zhuang2024tip} are straightforward, making segmentation essential. A dominant approach is 2D mask-guided segmentation~\cite{chen2024gaussianeditor, jiang2024vr,liu2024infusion, mirzaei2024reffusion, zhuang2024tip}, which lifts semantics from multi-view images using models like SAM~\cite{kirillov2023segment} and CLIP~\cite{radford2021learning}. 
Methods such as Gaussian Grouping~\cite{ye2024gaussian}, SAGA~\cite{cen2025segment}, and LangSplat~\cite{qin2024langsplat} embed semantic features into Gaussians and align them with 2D masks, but often suffer from 2D–3D inconsistencies and require multi-view training, limiting flexible and repeated editing (Fig.~\ref{fig:selection} (a)).
Interactive methods aim to improve usability, such as GaussianCut~\cite{jain2024gaussiancut} and ArtisanGS~\cite{tsang2026artisangs}, but still depend on 2D supervision and operate at coarse object levels. Similarly, tools like SuperSplat~\cite{supersplat2024} use screen-space selection via ray casting, which suffers from depth ambiguity (Fig.~\ref{fig:selection} (b)).
In contrast, our Super-Gaussian operates directly in 3D by clustering feature-aware Gaussian groups without 2D masks or predefined semantics, enabling intuitive region-level selection (Fig.~\ref{fig:selection} (c)). Compared to GaussianShopVR (GSVR)~\cite{shen2025gaussianshopvr}, which relies on point-by-point selection, our method introduces hierarchical select-and-refine interactions, reducing user effort and enabling efficient ROI specification.

\vspace{-0.075in}
\section{Preliminary}
\label{sec:preliminary}
3DGS~\cite{kerbl2023gs} represents a scene using a collection of explicit 3D Gaussian primitives, each associated with a set of differentiable attributes that can be rendered efficiently through tile-based rasterization.
Given the position attribute $\bm{\mu}$ of a Gaussian primitive, it is defined as
\vspace{-0.5em} 
\begin{equation}
	g(\mathbf{x})=e^{-\tfrac{1}{2}(\mathbf{x}-\bm{\mu})^\top \Sigma^{-1}(\mathbf{x}-\bm{\mu})},
\end{equation}
where $\mathbf{x}$ is an arbitrary 3D position and $\Sigma$ denotes the covariance matrix of the Gaussian primitive.
To ensure positive semi-definiteness during optimization, $\Sigma$ is formulated by two scaling and rotation matrices that are parameterized by a scaling factor $\mathbf{s}$ and a quaternion rotation vector $\mathbf{q}$, respectively.
During rendering, a 3D Gaussian primitive $g(\mathbf{x})$ is first projected onto a 2D Gaussian $g'(\mathbf{x'})$ on the image plane following the EWA volume splatting~\cite{Zwicker-VIS01}, and the pixel color $\mathbf{C}(\mathbf{x'})$ at 2D position $\mathbf{x'}$ is then computed via blending $M$ depth-sorted 2D Gaussians using their color $\mathbf{c}$ and opacity $o$ attributes, i.e.,
\vspace{-0.5em} 
\begin{equation}
    \mathbf{C}(\mathbf{x'}) = \sum_{i \in M} c_i \alpha_i \prod_{j=1}^{i-1} (1 - \alpha_j), \ \alpha_i = o_i g_i'(\mathbf{x'}),
\end{equation}
where $\mathbf{c}$ is modeled by a set of spherical harmonics (SH) parameters.
With the differentiable rasterizer, all attribute values are jointly optimized via training-view reconstruction.

However, directly applying 3DGS to fit VolVis scenes bakes appearance information (color and lighting) into the Gaussian primitive representation, making them uneditable during rendering and severely limiting interactive volume exploration in VR.
To address this, several studies~\cite{tang2025ivrgs, tang2025texgs, ai2025nli4volvis} in VolVis introduce editable Gaussian primitives that decouple color and lighting information during training, enabling color, opacity, and lighting editing at inference time.
Specifically, each editable Gaussian, in addition to the existing geometry-related attributes $\{\bm{\mu}, \mathbf{q}, \mathbf{s}, o\}$ for vanilla 3DGS, is equipped with a set of differentiable shading-related attributes $\{\Delta\mathbf{c}, \mathbf{n}, k^a, k^d, k^s, \beta \}$. 
Among them, $\Delta\mathbf{c}$ denotes the offset color, which, together with a learnable palette color $\mathbf{c}_p$ shared by all Gaussians within a basic scene, represents the diffuse color of each Gaussian. 
The attribute $\mathbf{n}$ encodes the normal direction, while $k^a$, $k^d$, $k^s$, and $\beta$ correspond to the ambient, diffuse, specular, and shininess coefficients in the Blinn-Phong shading model~\cite{Blinn-Phong}.
The final shaded color $\mathbf{c}$ of each Gaussian is computed as
\vspace{-0.5em} 
\begin{equation}
	\mathbf{c} = k^a \mathbf{c}_d + k^d \max(\mathbf{n} \cdot \bm{l}, 0)\, \mathbf{c}_d + k^s \max(\mathbf{n} \cdot \bm{h}, 0)^{\beta},
\end{equation}
where $\mathbf{c}_d = \mathbf{c}_p + \Delta\mathbf{c}$ is the diffuse color, $\bm{l}$ is the light direction, and $\bm{h}$ is the halfway vector between the view and light directions.
By optimizing these shading attributes during training, various editing operations can be achieved at inference time: adjusting the shared palette color $\mathbf{c}_p$ changes the color of the basic scene, scaling $o$ modifies its opacity, and altering the light source position or scaling $k^a$, $k^d$, $k^s$, and $\beta$ enables intuitive control over the lighting direction and intensity. 

Furthermore, the editable Gaussian models exhibit strong composability: by aggregating parameters from multiple basic models into a single composed model, all basic scenes can be jointly rendered without additional optimization, enabling flexible exploration of the complete VolVis scene under varying predefined TFs.


\vspace{-0.075in}
\section{Method}

We formulate volumetric ROI selection in VR as interaction over feature-aware Gaussian groups, termed \textit{Super-Gaussian}, and design a hierarchical select-and-refine workflow that couples graph-based propagation with cluster-level and primitive-level correction.

\vspace{-0.075in}
\subsection{Super-Gaussian Framework}

As shown in Fig.~\ref{fig:teaser}, our Super-Gaussian framework enables editable Gaussian-based VolVis in immersive VR environments, integrating spatial interaction with NLI within a unified system architecture. The system has three main components: (1) a {\bf VR client} responsible for immersive rendering and spatial interaction, (2) a {\bf NLI agent} that interprets natural language instructions and coordinates interaction logic~\cite{ai2025nli4volvis}, and (3) a {\bf backend server} that manages Gaussian selection, semantic labeling, and open-vocabulary query functions. By combining editable Gaussian representations, immersive interaction, and open-vocabulary language interfaces, the framework enables a tightly integrated VPA loop~\cite{liu2024-vpa} for exploring complex volumetric datasets.

\textbf{Pipeline overview.}
The visualization pipeline begins by loading a set of editable Gaussian models trained from multi-view images (Fig.~\ref{fig:teaser} (a)) into the VR environment, which are rendered together as a unified VolVis using Gaussian splatting. Within this immersive environment, users can interactively explore the VolVis scene and define ROIs at a finer granularity than the original base models (Fig.~\ref{fig:teaser} (b)). They can select arbitrary volumetric regions from the base units and convert them into editable units (Fig.~\ref{fig:teaser} (c)), which are treated as user-defined ROI represented as interactive primitives for subsequent editing and NLI-based manipulation.
Both the base units and user-selected editable units can be modified by adjusting their visualization attributes, including color, opacity, and lighting parameters. These editing operations can be performed either via GUI controls in the VR environment or via natural-language commands processed by the NLI agent. 
%
%
The NLI agent (Fig.~\ref{fig:teaser} (d)) extends the hierarchical multi-agent architecture of NLI4VolVis~\cite{ai2025nli4volvis}, consisting of a core agent for high-level reasoning, an intent parsing agent for interpreting user commands, and a function calling agent for open vocabulary querying,
3D scene editing and text labeling. Building on this structure, we further support spatial interaction and user-defined ROI manipulation in VR, enabling NLI for deeper scene understanding of user-selected regions (Fig.~\ref{fig:teaser} (e)).
%
To support NLI, a dedicated capture camera provides the rendered scene as perceptual input to the core agent, mitigating UI occlusion and gaze misalignment in VR.
Finally, NLI is enabled through a speech interface that integrates speech-to-text (STT) and text-to-speech (TTS) pipelines. This allows users to issue commands and receive system responses through voice interaction while simultaneously performing spatial manipulation in VR. 
In addition, each editable unit can be assigned a user-defined textual label (Fig.~\ref{fig:teaser} (f)). These labels extend the interaction space by enabling semantic referencing through natural language (Fig.~\ref{fig:teaser} (g)), allowing users to manipulate previously defined regions via open-vocabulary queries.




\textbf{Rendering editable 3DGS.}
To render our editable Gaussian representation in VR, we extend an open-source Unity-based Gaussian splatting renderer~\cite{UnityGaussianSplatting_2025}, a Unity-native 3DGS implementation that performs GPU-based preprocessing and depth sorting, and rasterizes each Gaussian as a procedurally generated screen-space quad using HLSL and compute shaders. 
Built on this VR rendering framework, our system preserves the baseline GPU sorting and splat compositing pipeline, while extending the Gaussian representation to support appearance-level editability. Specifically, each Gaussian is augmented with additional editable attributes, including a surface normal, opacity, an offset color, and ambient, diffuse, specular, and shininess coefficients, which are reorganized and packed into GPU-readable buffers during asset import. At runtime, these attributes are decoded during view preprocessing and combined with user-controlled palette and lighting parameters to reconstruct the final splat appearance, enabling interactive recoloring and relighting directly in VR. To support interactive selection and editable units, we maintain persistent per-selection masks as bitwise GPU buffers. During rendering, each splat is assigned a mark based on its editable unit membership, which is used to apply unit-specific color and opacity overlays in the fragment stage. This design allows the renderer to support editable Gaussian visualization with persistent, appearance-aware unit highlighting and interaction while retaining real-time VR performance.
\sm{For training editable 3DGS, we adopt a two-stage training scheme. In the first stage, we train a vanilla 3DGS model~\cite{kerbl2023gs} for 30,000 iterations to capture the scene geometry.
In the second stage, we extend the vanilla 3DGS model to include shading-related attributes (refer to Section~\ref{sec:preliminary}) and optimize them for an additional 10,000 iterations.
Both stages combine L1, SSIM, and regularization losses~\cite{tang2025ivrgs} to achieve high-fidelity reconstruction.
}

\textbf{User interface.}
The user interface is divided into three main components, as shown in Appendix Fig.~\ref{fig:appendix-ui} (a): {\bf scene control panel}, {\bf 3D prompt panel}, and {\bf NLI interface}. 
The scene control panel provides direct controls for editing Gaussian-based visualizations. Users can select which base units and their associated TF to manipulate, and previously defined editable units appear under the selected list for further editing. For each selected unit, visualization attributes such as color, opacity, lighting parameters, and viewpoint can be adjusted.
Appendix Fig.~\ref{fig:appendix-ui} (b) shows the selection panel used for interactive region selection. Users can adjust the brush width and manage the selection state through reset and save operations. 
The system supports three complementary selection modes. The {\em random walk (RW) mode} automatically selects super Gaussians based on user-provided foreground and background seeds. 
The {\em cluster mode} enables manual selection of super Gaussians, while the {\em point mode} supports individual Gaussian selection for fine-grained cluster boundary refinement.
These modes can be combined in the select-and-refine workflow to progressively refine the target region. Selected regions are highlighted with a configurable selection color to distinguish them from the current volumetric rendering. Users can also temporarily modify the color and opacity of selected regions before committing them as editable units.
Under the selection controls, the new text-label field lets users assign semantic labels to the selected region. Once saved, the labeled editable unit is automatically added to the selected list in the scene control panel, enabling subsequent editing through both GUI controls and NLI.
The NLI interface supports speech-based interaction via an STT pipeline and displays the conversation history in text. This interface allows users to review issued commands, inspect agent actions, and view query responses generated by the system.
In addition to the panel-based interface, users can directly manipulate the volumetric scene using VR controllers, including translation, scaling, and rotation. Rotation can also be controlled via the controller's thumbstick. Brush interaction is performed using trigger inputs from both hands, allowing users to select and deselect through continuous 3D brush strokes.




\vspace{-0.075in}
\subsection{Interactive ROI Selection}


\subsubsection{Super-Gaussian}
\label{sec:supergs}
\begin{figure}[t]
\centering
\includegraphics[width=\linewidth]{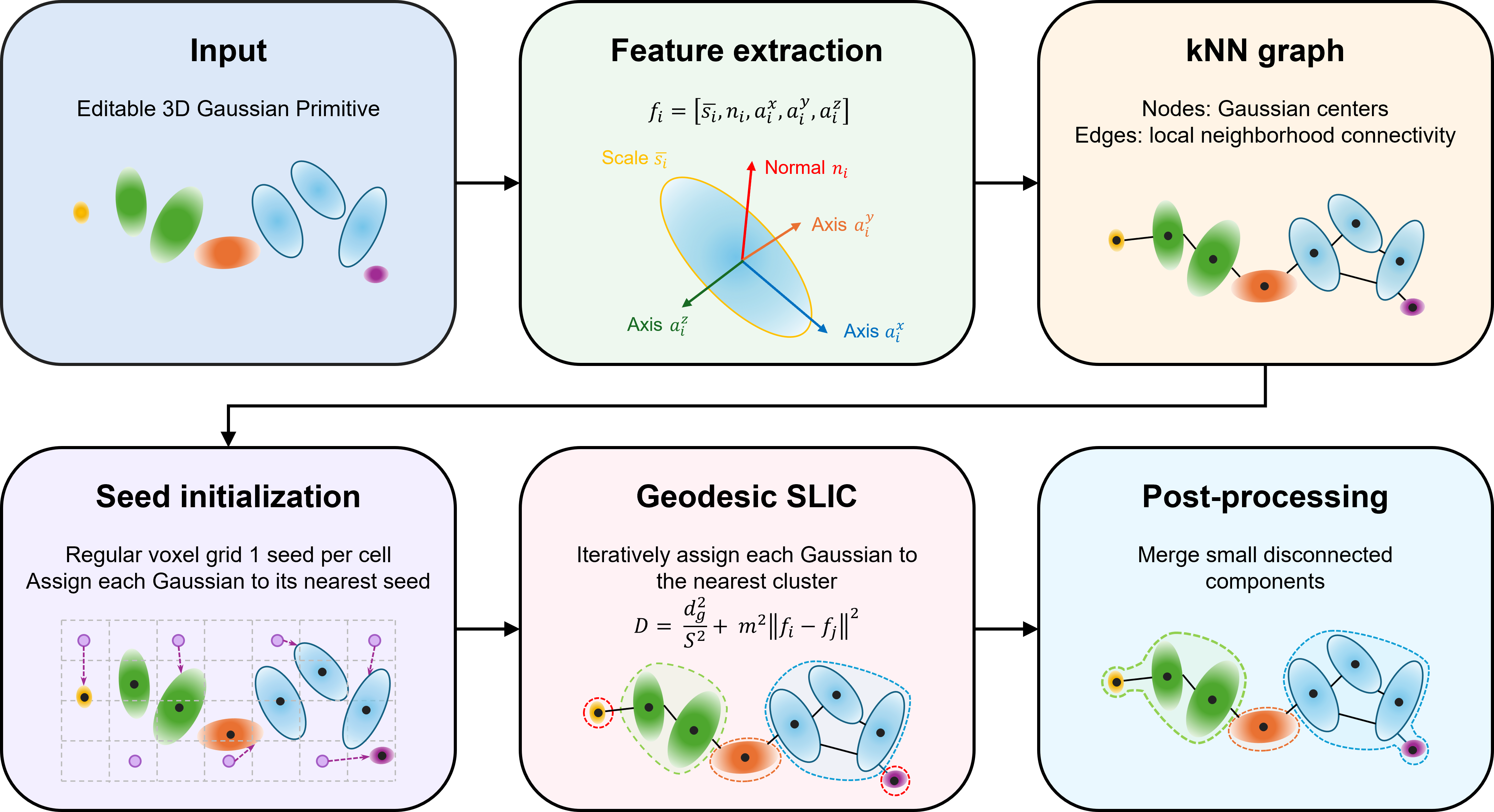}
\vspace{-0.175in}
\caption{\sm{
Block diagram of the Super-Gaussian clustering pipeline.}}
\label{fig:clustering}
\vspace{-0.1in}
\end{figure}

Inspired by superpixel methods in image processing, which group pixels into perceptually meaningful regions using simple linear iterative clustering (SLIC)~\cite{achanta2012slic}, we introduce \textit{Super-Gaussian} as a structure-aware cluster of Gaussian primitives. 
Similar to how superpixels provide a higher-level representation for images, \textit{Super-Gaussian} serves as a higher-level interaction unit for ROI selection. 
While SLIC groups pixels by color and 2D spatial position, we extend this formulation to 3D Gaussian representations by incorporating 3D spatial positions and intrinsic Gaussian features, and adopt a geodesic SLIC formulation to better preserve structural continuity.
\sm{Fig.~\ref{fig:clustering} provides a high-level overview of the Super-Gaussian construction pipeline.}

In editable Gaussian representations, various attributes are available for each Gaussian, including appearance and lighting parameters. However, not all of these attributes are suitable for defining structural similarity. 
As shown in Appendix Fig.~\ref{fig:feature}, geometric attributes (e.g., mean scale, normal, principal axes) exhibit coherent spatial patterns aligned with object structures, with significant variation near edges or structural transitions, whereas appearance-related features (e.g., offset color, opacity) show minimal variation or resemble random noise.
For example, offset color $\Delta\mathbf{c}$ 
reflects appearance rather than intrinsic geometry. 
Similarly, opacity values can be noisy due to rendering optimization, and lighting-related attributes often exhibit little variation within a base model. 
In contrast, the mean scale reflects the local spatial extent of each Gaussian and often varies near thin structures and structural transitions, while the normal and principal axes capture local orientation patterns.
To obtain stable structure-aware clusters, we therefore construct the feature representation using only geometric attributes derived from the Gaussian parameters.


Based on this observation, each Gaussian $g_i$ is represented by a compact feature vector that captures both local geometric scale and orientation for clustering. 
The feature vector is defined as
\vspace{-0.5em} 
\begin{equation}
\mathbf{f}_i =
\left[
\bar{s}_i,\,
\mathbf{n}_i,\,
\mathbf{a}_i^{x},\,
\mathbf{a}_i^{y},\,
\mathbf{a}_i^{z}
\right]
\in \mathbb{R}^{13},
\end{equation}
where $\bar{s}_i = (s_{i,x} + s_{i,y} + s_{i,z})/3$ denotes the mean scaling of the Gaussian, computed as the average of its scaling vector $\mathbf{s}_i = (s_{i,x}, s_{i,y}, s_{i,z})$,
$\mathbf{n}_i$ represents the normal direction, and 
$\mathbf{a}_i^{x}$, $\mathbf{a}_i^{y}$, and $\mathbf{a}_i^{z}$ correspond to the local principal axes derived from the Gaussian quaternion rotation vector $\mathbf{q}$ along three directions. 
$\bar{s}_i$ captures the overall size of $i$-th Gaussian, while $\mathbf{n}_i$ and $[ \mathbf{a}_i^{x},\mathbf{a}_i^{y}, \mathbf{a}_i^{z} ]$ encode the local directional structure. 
This feature representation allows the clustering process to group Gaussians that are not only spatially nearby but also geometrically consistent.

\begin{figure}[t]
\centering
\includegraphics[width=\linewidth]{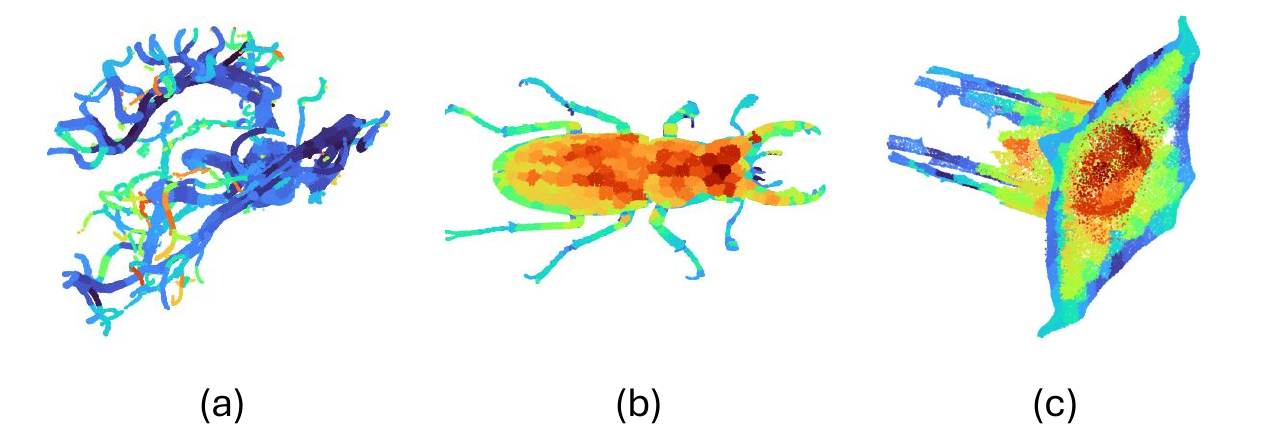}
\vspace{-0.275in}
\caption{Super-Gaussian for (a) \texttt{aneurysm}, (b) \texttt{beetle}, and (c) \texttt{ionization}, color-mapped from cluster-level features by projecting each feature vector onto the first PCA axis.
}
\label{fig:supergs}
\vspace{-0.1in}
\end{figure}

To preserve structural continuity, clustering is performed using graph-based geodesic distance rather than solely Euclidean distance.
Following classical manifold learning methods that approximate geodesic distances via shortest paths on $k$NN graphs for unstructured point clouds \cite{tenenbaum2000global, klein2004proximity}, we construct a point-wise $k$NN graph over individual Gaussian centers and compute geodesic distances as shortest-path distances on this graph.
Cluster seeds are initialized on a regular voxel grid, placing one seed per cell to ensure uniform spatial coverage. 
Starting from these seeds, the $k$-means clustering algorithm iteratively assigns each Gaussian $g_i$ to a cluster center $j$ that minimizes a combined distance metric defined as
\vspace{-0.5em} 
\begin{equation}
D(i,j)
=
\frac{d_g(i,j)^2}{S^2}
+
m^2 \left\|\mathbf{f}_i - \mathbf{f}_j \right\|^2,
\end{equation}
where $d_g(i,j)$ denotes the geodesic distance between Gaussian $g_i$ and cluster center Gaussian $g_j$(Gaussian closest to the cluster center location), 
$S$ denotes the voxel cell size used for seed initialization,
$\mathbf f_i$ and $\mathbf f_j$ denote the feature vectors of Gaussian $g_i$ and cluster center Gaussian $g_j$, respectively, and 
$m$ balances the influence between spatial proximity and feature similarity. 
Specifically, geodesic distances are computed on a point-wise $k$NN graph over Gaussian centers as
\vspace{-0.5em} 
\begin{equation}
d_g(i,j)
=
\min_{p \in \mathcal{P}(i,j)}
\sum_{(u,v)\in p}
\left\| \bm{\mu}_u - \bm{\mu}_v \right\|,
\end{equation}
where $\mathcal{P}(i,j)$ represents all possible paths connecting the two Gaussians in the graph, and shortest paths are found via Dijkstra's algorithm.
During clustering, Gaussian assignments and cluster centers are iteratively updated in both spatial and feature spaces.
After convergence, small disconnected components are merged into neighboring clusters to reduce topological noise. 
Consequently, as shown in Fig.~\ref{fig:supergs}, each Super-Gaussian forms a coherent local unit that better follows volumetric structures than purely spatial clustering. 
Note that super Gaussians are precomputed for each input base scene, enabling efficient interactive selection and RW operations in response to user input. 
In practice, the number of super Gaussians per base scene is kept below 1,000, providing a compact graph-level representation that significantly reduces the complexity of subsequent selection and propagation operations.



\vspace{-0.05in}
\subsubsection{Selection Mode}

Based on precomputed super Gaussians, we provide three selection modes: \textit{RW}, \textit{cluster}, and \textit{point}.
\textbf{RW mode:}
In RW mode, brush strokes are densely sampled in 3D, mapped to the nearest Gaussian primitives, and converted to their corresponding Super-Gaussian cluster IDs, which are accumulated as foreground or background seeds.
When no background seeds are provided, a set of spatially distant foreground clusters is automatically designated as background seeds to ensure stable computation. 
Selection is then propagated over a precomputed Super-Gaussian graph $G=(V,E,W)$ using an RW algorithm~\cite{pearson1905problem}, where each node in $V$ represents a Super-Gaussian cluster, and edge weights in $W$ encode geodesic proximity and feature affinity between neighboring clusters.
Specifically, the graph is constructed by first building a cluster-level $k$NN over Super-Gaussian centroids using Euclidean distances,
and then computing geodesic distances $g_{uv}$ between cluster centroids via Dijkstra's algorithm on this adjacency graph.
Cluster features are $z$-score normalized, and the edge weight between neighboring clusters $u$ and $v$ is defined as
\vspace{-0.5em} 
\begin{equation}
w_{uv} = 0.9 \exp \big( - ( \frac{g_{uv}}{\sigma_s} )^2 \big)
+ 0.1 \exp \big( - ( \frac{\lVert \tilde{\mathbf{f}}_u - \tilde{\mathbf{f}}_v \rVert}{\sigma_f} )^2 \big)
\end{equation}
where ${\sigma_s}$ and ${\sigma_f}$ control the sensitivity to spatial and feature distances, respectively. 
Given the graph Laplacian $L=D-W$, the foreground probabilities for unlabeled nodes are obtained by solving
\vspace{-0.5em} 
\begin{equation}
L_{uu}\bm{\mu}_u = -L_{ul}\bm{\mu}_l .
\end{equation}
where $\bm{\mu}_l$ denotes the fixed labels of the foreground and background seeds, and ${\mu}_u$ denotes the unknown probabilities of the unlabeled nodes.
The resulting probabilities are thresholded at $0.5$ to determine the selected Super-Gaussian regions.
Since the RW is solved directly on the Super-Gaussian graph rather than on individual Gaussian primitives, the system involves at most a few hundred to~1,000 nodes per scene, enabling near-instantaneous propagation (\textasciitilde 10ms) even for large volumetric datasets.
%
%
\textbf{Cluster mode:}
In this mode, users can directly toggle multiple super Gaussians intersected by a brush stroke. Clusters can be explicitly added to or removed from the selection set without graph propagation.
%
\textbf{Point mode:}
This mode operates directly on Gaussian primitives. 
%
The Gaussians intersected by the brush tube (see Sec.~\ref{sec:brush} for details) 
are selected individually, enabling fine-grained editing of 
cluster boundaries. 

{\bf Select-and-refine workflow.}
The above three selection modes are designed to operate in a select-and-refine manner. RW mode provides a fast initial estimate from sparse foreground/background cues. Cluster mode then allows users to coarsely correct over-selected or under-selected regions at the Super-Gaussian level. Finally, Point mode enables fine-grained boundary refinement and detail correction at the primitive level. Since the system maintains accumulated foreground/background edits across modes, these interactions are mutually complementary rather than isolated operations.



\vspace{-0.05in}
\subsubsection{Brush Interaction}
\label{sec:brush}

Brush interaction serves different roles depending on the selection mode. In RW mode, it specifies foreground and background seeds, whereas in cluster and point modes, it selects super Gaussians or individual Gaussians. 
To support this interaction, we use two types of brushes: a green brush for foreground specification (i.e., adding to the selection) and a red brush for background specification (i.e., removing from the selection).
Regardless of the type, brush strokes are first densified and interpreted as 3D tubular regions to identify nearby Gaussian primitives.
Brush input is represented as one or more brush strokes in normalized volume coordinates.
Each stroke is defined as a polyline composed of ordered control points
$\{\mathbf{p}_0, \mathbf{p}_1, \dots, \mathbf{p}_n\}$.
A continuous stroke segment between two points is parameterized as
\vspace{-0.5em} 
\begin{equation}
\mathbf{p}(t) = (1-t)\mathbf{p}_i + t\mathbf{p}_{i+1}, \qquad t \in [0,1].
\end{equation}
To avoid gaps caused by sparse sampling during fast hand motion, each stroke is densified by inserting additional samples along the polyline.
Given $N$ samples along a segment, the sampled points are
\vspace{-0.5em} 
\begin{equation}
t_j = \frac{j}{N}, \qquad j = 0,\dots,N.
\end{equation}
The sampled points are mapped from normalized coordinates to the volume coordinate system using the bounding box. 
Each stroke is interpreted as a tubular region with radius $r$. 
Let $\bm{\mu}_i$ denote the center of Gaussian $g_i$ and $\mathbf{p}_j$ a sampled brush point. 
Gaussians intersecting the brush are identified as
\vspace{-0.5em} 
\begin{equation}
\mathcal{S}
=
\left\{
g_i \;\middle|\;
\exists \mathbf{p}_j \ \text{s.t.} \
\| \bm{\mu}_i - \mathbf{p}_j\| < r
\right\},
\end{equation}
where $\mathcal{S}$ represents the set of selected Gaussians.
The resulting selected Gaussians are used differently depending on the selection mode.
For RW and cluster modes, the identified Gaussians are
converted to their corresponding Super-Gaussian cluster IDs and used
either as foreground/background seeds for graph-based propagation or as explicit cluster-level toggles.
In point mode, the selected Gaussian primitives are selected directly.

\begin{figure*}[t]
\centering
\includegraphics[width=0.925\linewidth]{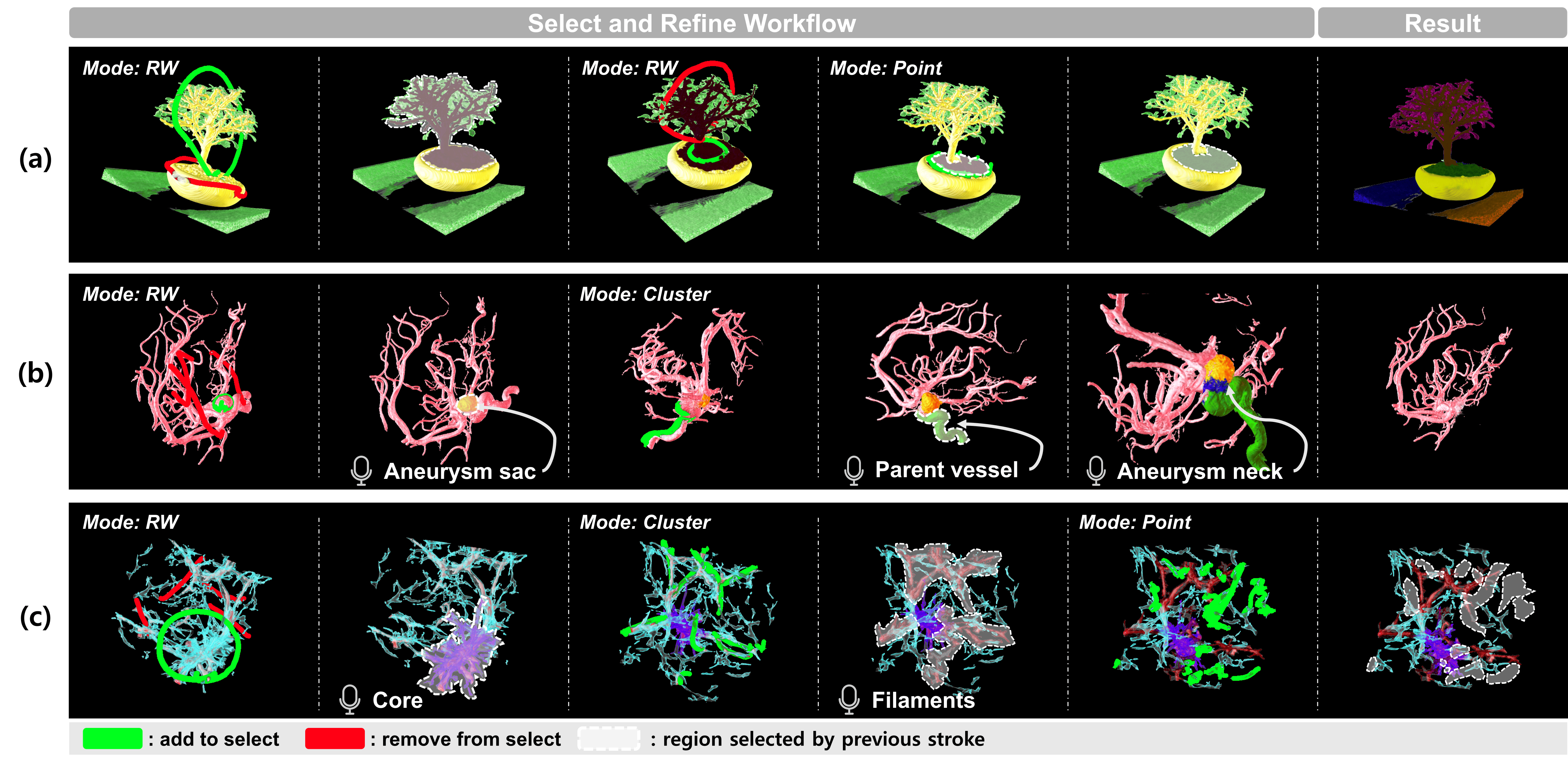}
\vspace{-0.15in}
\caption{The select-and-refine workflow with text labeling example of (a) \texttt{bonsai}, (b) \texttt{aneurysm}, and (c) \texttt{Nyx} datasets, respectively. The upper-left corner indicates the user selection mode for the brush input, and the mic icon with text indicates the newly added text label for the selected part. \sm{White dashed outlines highlight modifications from the previous step.}}
\label{fig:bonsai_select_refine}
\vspace{-0.1in}
\end{figure*}

\vspace{-0.05in}
\subsection{NLI-guided Selected Region Labeling}

Our framework enables semantic labeling of user-selected regions before storing them as editable units to support NLI.
Text labeling can be initiated either directly through the user interface or via a natural language command. In the latter case, the VPA loop is engaged: the NLI agent perceives the currently selected region, the intent-parsing module determines whether the command corresponds to a labeling action, and the function-calling agent invokes the labeling function accordingly.
When a user selects an ROI and provides a text label through either modality, the system retrieves the Gaussian indices of the corresponding base model. It renders view-dependent images of the selected region from a selected subset of Gaussians.
%
To obtain a representative visual description of the selected region, we adopt the entropy-guided view selection strategy from NLI4VolVis~\cite{ai2025nli4volvis}, applying it to user-selected Gaussian subsets rather than full scenes as in the original work.
Specifically, the information content of each rendered frame is evaluated using an opacity entropy metric. 
Given the opacity channel $p$, the entropy of a frame is computed as
\vspace{-0.5em} 
\begin{equation}
H = -\sum p \log p.
\end{equation}
Frames with higher entropy typically contain richer structural information and are therefore preferred. 
We select the top-$K$ frames with the highest entropy scores and compute their CLIP image embeddings. 
The image embedding for the selected region is obtained by averaging the embeddings of the selected frames
\vspace{-0.5em} 
\begin{equation}
\mathbf{e}_I = \frac{1}{K}\sum_{i=1}^{K} f(I_i),
\end{equation}
where $f(\cdot)$ denotes the CLIP image encoder and $I_i$ represents the selected frames. 
The resulting embedding is normalized to obtain a stable semantic representation.
In parallel, the user-provided text label is encoded using the CLIP text encoder to produce a text embedding $\mathbf{e}_T$. 
Following the representation strategy adopted in NLI4VolVis~\cite{ai2025nli4volvis}, the final semantic representation is obtained by combining the image and text embeddings
\vspace{-0.5em} 
\begin{equation}
\mathbf{e} = \frac{\mathbf{e}_I + \mathbf{e}_T}{2}.
\end{equation}
The fused semantic embedding $\mathbf{e}$ is passed to an open-vocabulary query function, enabling users to reference the selected region in natural language and to perform subsequent scene-editing operations via semantic commands. On-the-fly text labeling takes 10-15 seconds, depending on network latency. This process runs asynchronously on the server, so users can continue exploring and editing the volumetric scene in VR while the CLIP-based labeling is computed.







\vspace{-0.05in}
\section{Case Studies}

We present four case studies to demonstrate how our framework extends NLI for VolVis through Super-Gaussian selection and immersive interaction in VR. While NLI4VolVis~\cite{ai2025nli4volvis} has already shown the usefulness of NLI for data visualization, existing approaches typically rely on global commands or pre-segmented structures. Our system enables users to interactively define ROIs via spatial selection and combine them with natural language queries, forming an integrated VPA loop.
Through four case studies across simulation, natural-object, and medical datasets, we highlight how integrating partial ROI selection with NLI extends existing NLI-driven data exploration toward richer analysis and presentation workflows.
The case studies show three key capabilities of our framework: (1) spatially guided interpretation of volumetric structures through partial Gaussian selection, (2) iterative perceptual exploration through the integration of selection and natural language queries, and (3) flexible communication and presentation of volumetric structures through interactive region specification.

\vspace{-0.075in}
\subsection{\texttt{Bonsai}}

We evaluate the usefulness of our framework's select-and-refine workflow using the \texttt{bonsai} dataset. This dataset is known to be difficult to separate using a simple 1D TF, because semantically different structures often share similar intensity values and are therefore mapped to the same visual appearance. Through our VR-based selection, users can intuitively distinguish regions with similar visual or semantic characteristics and assign them different colors. Starting from base models trained with two TFs, we demonstrate how users can progressively refine the visualization by separating and coloring structures such as grass, branches, leaves, and rocks. As shown in Fig.~\ref{fig:bonsai_select_refine} (a), we first separate the pot region from the surrounding structures using the RW selection. To initialize RW, users roughly indicate foreground and background regions using a circular brush. The brush strokes do not need to provide precise seeds; instead, they only need to roughly indicate the ROIs, which is sufficient for RW to separate the pot from the remaining structures. After this coarse separation, we further refine the selection. In the pot and gravel regions, which have irregular and detailed boundaries, users perform additional manual selection to isolate the necessary areas more precisely. Finally, structures such as leaves and grass, which have similar intensity values, are separated by spatial distance. Because these regions are clearly separated in space, a single RW selection can capture the entire leaf structure, while the grass regions can also be selected and colored accordingly. This step-by-step refinement demonstrates how our workflow enables intuitive separation and labeling of semantically meaningful structures even when they share similar intensity values.


\vspace{-0.1in}
\subsection{\texttt{Aneurysm}}

\begin{figure}[t]
\vspace{-0.1in}
\centering
\includegraphics[width=\linewidth]{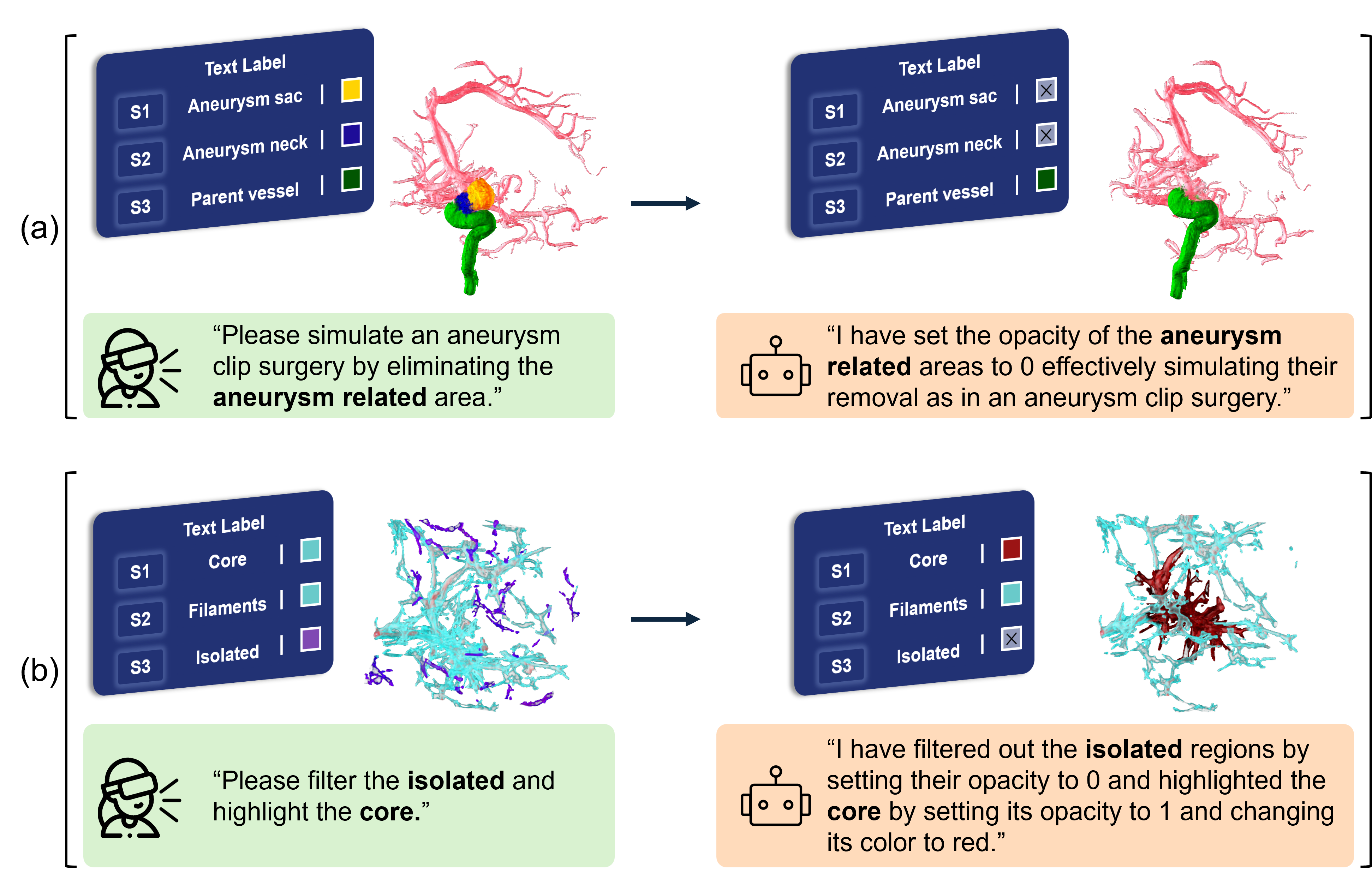}
\vspace{-0.25in}
\caption{
Case study with the \texttt{aneurysm} and \texttt{Nyx} datasets demonstrating the use of NLI with user-defined text labels. Each schematic VR panel summarizes the text labels and colors of editable units.}
\label{fig:nli1}
\vspace{-0.1in}
\end{figure}

Medical datasets, such as the \texttt{aneurysm} dataset, can benefit from VR-based selection and interactive text labeling. This dataset captures vascular structures with aneurysms, in which identifying clinically relevant regions is important for analysis and surgical planning. VR VolVis has often been used to explore such datasets, for example, to inspect anatomical structures or simulate surgical procedures before an operation.
As shown in Fig.~\ref{fig:bonsai_select_refine} (b), we first use selection to identify the aneurysm and decompose it into three clinically meaningful components: the \emph{aneurysm sac}, the \emph{parent vessel}, and the \emph{aneurysm neck}. While navigating the volume and rotating the view, users interactively select the connected vascular structures and assign corresponding labels. In particular, the neck region, which connects the sac to the parent vessel, is of special interest because it is typically the target for surgical clipping, where a clip is placed across the neck to isolate the sac from blood flow while preserving the parent vessel.
Through natural language instructions, users can simultaneously manipulate all selected units associated with this label, simulating a surgical planning scenario (Fig.~\ref{fig:nli1} (a)). For instance, users can reduce the opacity of the aneurysm-related area to simulate clipping or removal, enabling clearer visualization of the remaining vessel structures.
In addition, users can examine the internal structure in greater detail and confirm the spatial relationship between the aneurysm and the surrounding vessels.
%
%


\vspace{-0.075in}
\subsection{\texttt{Nyx}}

For the \texttt{Nyx} dataset, we demonstrate how users can explore and interpret complex cosmological structures through a step-by-step classification and labeling process. We first classify visible structures based on their geometric characteristics  (Fig.~\ref{fig:bonsai_select_refine} (c)), separating a dense central core, elongated filament-like branches extending from the core, and smaller isolated fragments distributed around the structure. We then assign semantic labels to these regions, identifying them as the core, filaments, and isolated structures. This process reveals meaningful components of the cosmic web that are otherwise difficult to distinguish in semi-transparent volume renderings.
Importantly, this example illustrates that our approach enables users to progressively interpret datasets without clear boundaries or predefined labels through interactive classification and annotation. Even spatially disconnected or visually ambiguous regions (e.g., isolated fragments) can be grouped under a single semantic label. Using NLI, users can then easily filter and manipulate these grouped regions, enabling efficient editing of complex simulation data (Fig.~\ref{fig:nli1} (b)).


\vspace{-0.075in}
\subsection{\texttt{Chameleon}}

\begin{figure}[t]
\vspace{-0.1in}
\centering
\includegraphics[width=\linewidth]{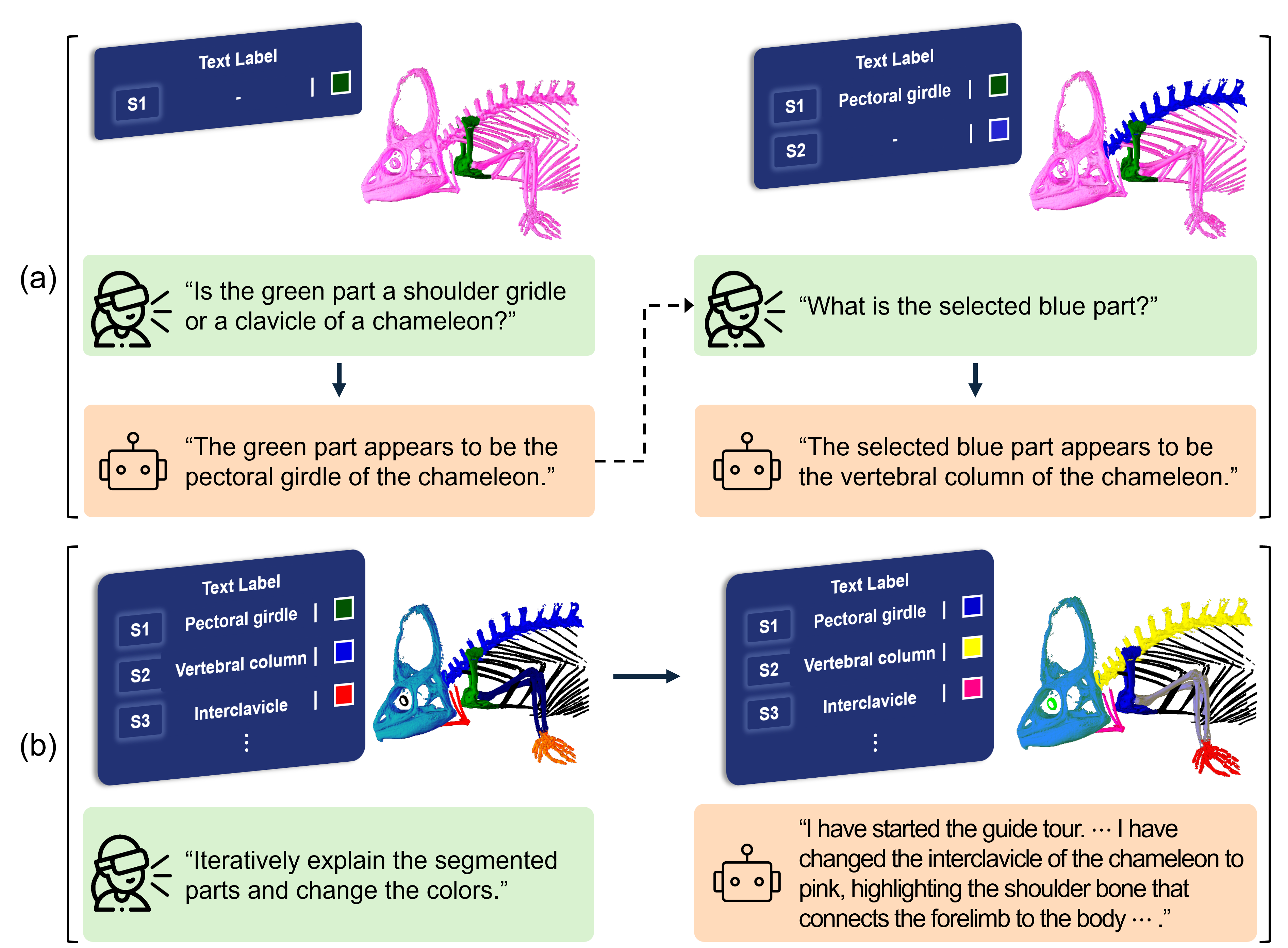}
\vspace{-0.25in}
\caption{Case study with the \texttt{chameleon} dataset demonstrating perception-guided NLI for selected regions.
(a) Question answering on selected regions and text labeling by the agent. (b) NLI agent performing an iterative guided tour based on the added text labels. Each schematic VR panel summarizes the text labels and colors of editable units.}
\label{fig:nli2}
\vspace{-0.1in}
\end{figure}

Our framework also enables anatomically meaningful labeling and supports more fine-grained analysis compared to conventional labeling approaches. Through a perception loop, users can interactively explore segmented regions and interpret their semantic meaning. For example, as illustrated in Fig.~\ref{fig:nli2} (a), a user may select a specific bone region of the chameleon and change its color to highlight the selected part. The user can then ask a question such as ``{\em Is the green part a shoulder girdle or a clavicle of a chameleon?}'' The system captures the current VR view with a capture camera, converts it to an image, and sends it along with the query to the LLM. Based on the visual context and the user query, the model can describe the selected region, for example: ``{\em The green part appears to be the pectoral girdle of the chameleon.}''
Similarly, users can query other selected regions, such as ``{\em What is the selected blue part?}'', and receive responses like ``{\em The selected blue part appears to be the vertebral column of the chameleon.}'' This interaction allows users to segment the dataset while simultaneously leveraging the LLM's background knowledge to interpret anatomical structures. Furthermore, users can review the overall segmentation results by issuing instructions (Fig.~\ref{fig:nli2} (b)) such as ``{\em Give me a guided tour,}'' enabling the system to summarize and explain the segmented structures across the dataset.

\vspace{-0.075in}
\section{Quantitative Evaluation}
\sm{We evaluate Super-Gaussian through objective technical measurements with three components: (1) a quantitative selection comparison against existing Gaussian-based selection methods, (2) rendering efficiency measurements in VR, and (3) an ablation study on geodesic SLIC clustering.}
All experiments were conducted on a desktop PC equipped with an AMD Ryzen 9 9950X 16-Core Processor, 128 GB of RAM, and an NVIDIA GeForce RTX 3090 GPU. For VR-based conditions, we used a Meta Quest 3 headset connected via a wired Link to the PC-based VR setup. The headset provides a resolution of 2064$\times$2208 per eye, a 110$^\circ$ field of view, and up to a 120 Hz refresh rate.

\vspace{-0.05in}
\subsection{Quantitative Selection Comparison}

\begin{figure}[t]
\centering
\includegraphics[width=0.9\linewidth]{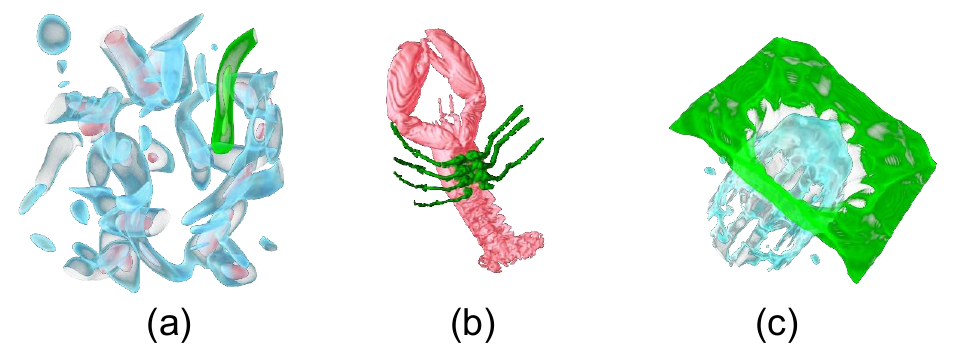}
\vspace{-0.125in}
\caption{Datasets used in the quantitative selection comparison: (a) \texttt{vortex}, (b) \texttt{lobster}, and (c) \texttt{ionization}. The target region is highlighted in green. 
}
\label{fig:userstudy_selection}
\vspace{-0.1in}
\end{figure}

\textbf{Datasets.}
To evaluate selection accuracy, interaction count, and completion time, we selected three datasets. Fig.~\ref{fig:userstudy_selection} highlights the target regions in green in all cases.
The \texttt{vortex} dataset features a transparent outer structure surrounding a dense inner core and is used to evaluate selection within transparent volumes with clear density separation. The \texttt{lobster} dataset includes surface-connected regions with thin leg structures, providing a test of performance on fine, articulated geometry. The \texttt{ionization} dataset contains a large, sparse target region with empty spaces, assessing selection in extended, low-density areas.

\textbf{Methods for comparison.}
We compared three selection methods: a 2D screen-space method (SuperSplat~\cite{supersplat2024}), a manual point-based VR method (GSVR~\cite{shen2025gaussianshopvr}), and our Super-Gaussian. 

\textbf{Protocol.}
The comparison was conducted by four trained author-operators under a fixed protocol. Each author-operator completed the same target-selection tasks for all method--dataset pairs. Before the recorded trials, the operators completed a familiarization session using the \textit{beetle} dataset to practice the input mappings and selection modes for each method.
These familiarization trials were not included in the reported results. The order of the methods was counterbalanced across operators to reduce ordering effects.
For each dataset, the operator first inspected the volume to identify the target region shown in Fig.~\ref{fig:userstudy_selection}. Timing began at the first selection action and ended when the operator confirmed the result using the ``Complete'' button. Operators were instructed to refine each selection until the visible selected region matched the target region as closely as possible.
In the VR conditions, GSVR and Super-Gaussian used the same controller-based interactions for volume manipulation, including rotation, translation, and scaling. Both methods used brush-based controller input to add and remove selections. For GSVR, operators manually selected Gaussian primitives, whereas Super-Gaussian allowed selection and refinement through RW, cluster, and point modes. In SuperSplat, operators used a mouse and keyboard to rotate, translate, scale, and refine the selection. Shortcut keys, including undo and navigation controls, were enabled when supported by the interface.






\begin{table}[t]
\centering
\caption{\sm{Quantitative selection comparison across datasets.} The completion time is in seconds. The best values are highlighted in bold.}
\label{tab:avg-all}
\vspace{-0.1in}
\begin{adjustbox}{width=\columnwidth}
\scriptsize
\setlength{\tabcolsep}{3pt}
\begin{tabular}{llrrrrrrr}
\toprule
 &  &  &  &  &  &  & int. & comp.\\
dataset & method & ACC & MCC & F1 & IoU & prec. & count & time\\
\midrule
& SuperSplat & 0.9937 & 0.9098 & 0.9065 & 0.8410 & 0.9751 & 19.50 & 119.43 \\
\texttt{vortex} & GSVR & 0.9898 & 0.8532 & 0.8488 & 0.7451 & 0.9741 & 12.50 & 138.30 \\
& Super-Gaussian & \textbf{0.9967} & \textbf{0.9551} & \textbf{0.9563} & \textbf{0.9176} & \textbf{0.9766} & \textbf{6.75} & \textbf{21.54} \\
\midrule

& SuperSplat & 0.9876 & 0.9507 & 0.9579 & 0.9200 & 0.9663 & 37.00 & 142.67 \\
\texttt{lobster} & GSVR & 0.9823 & 0.9284 & 0.9360 & 0.8822 & \textbf{0.9872} & 26.50 & 109.50 \\
& Super-Gaussian & \textbf{0.9927} & \textbf{0.9714} & \textbf{0.9754} & \textbf{0.9523} & 0.9809 & \textbf{25.50} & \textbf{94.18} \\
\midrule

& SuperSplat & 0.9826 & 0.9655 & 0.9838 & 0.9681 & 0.9982 & 11.25 & 85.44 \\
\texttt{ionization} & GSVR & 0.8132 & 0.6379 & 0.8222 & 0.7347 & 0.8822 & 22.75 & 91.88 \\
& Super-Gaussian & \textbf{0.9856} & \textbf{0.9713} & \textbf{0.9866} & \textbf{0.9736} & \textbf{0.9992} & \textbf{11.00} & \textbf{37.04} \\

\bottomrule
\end{tabular}
\end{adjustbox}
\vspace{-0.1in}
\end{table}

\textbf{Accuracy.}
To assess selection performance, we report multiple accuracy metrics, including overall accuracy (ACC), Matthews correlation coefficient (MCC), F1 score, intersection over union (IoU), and precision, as shown in Table~\ref{tab:avg-all}. 
These metrics are computed based on true positives (TP; correctly selected Gaussians), false positives (FP; incorrectly selected Gaussians), and false negatives (FN; target Gaussians that were not selected). In our setting, all Gaussians in each dataset are considered, with those in the target region treated as positives and those outside it as negatives.
While ACC and F1 capture overall correctness and the balance between precision and recall, they can be biased in cases with class imbalance, which is common in point cloud selection, where target regions occupy only a small portion of the data~\cite{yu2012efficient, yu2015cast,zhao2023metacast}. To address this, we additionally report MCC, which incorporates true negatives (TN; correctly unselected Gaussians) and provides a more balanced evaluation across all classes. IoU and precision are also included to provide complementary perspectives on the quality of overlap and the correctness of selection.

\textbf{Interaction efficiency.}
\sm{To quantify operational costs,} we measured interaction counts and completion times.
For interaction count, we defined a unified notion of a ``selection action'' across different methods. In 2D-based methods (e.g., brush, lasso, polygon, and rectangle selection), a single selection operation is counted as one interaction. However, different tools use different input patterns; for example, lasso and rectangle selections involve a click-and-drag followed by a release, while polygon selection requires multiple discrete clicks to form a region.
\sm{For VR-based methods (GSVR and Super-Gaussian), a single continuous brush stroke counts as a single interaction.}
Since these interaction modalities are inherently different, we additionally measure completion time, defined as the duration from the start of the first selection action to the completion of the final selection.
Furthermore, when comparing VR-based methods, we consider the variability within a single brush stroke. A single interaction may involve either a short or prolonged mid-air gesture while holding the controller trigger. To better capture this difference, we sample brush center points at regular intervals during each stroke and compute the average number of sampled points per stroke. This provides a finer-grained measure of interaction \sm{efficiency} for VR-based techniques.
\sm{We report these metrics as operational measures for comparing method behavior, rather than as subjective measures of user effort or workload.}

%
%
%
\textbf{Results.}
The results show that Super-Gaussian consistently outperforms baseline methods (SuperSplat and GSVR) in both \sm{accuracy and interaction cost.} 
%
%
Table~\ref{tab:avg-all} demonstrates that Super-Gaussian achieves the highest overall accuracy across datasets, though performance varies depending on dataset characteristics. For datasets with clearly defined boundaries (e.g., \texttt{vortex} and \texttt{ionization}), SuperSplat and GSVR can sometimes achieve comparable or slightly higher accuracy, likely due to effective 2D-based refinement interactions. In contrast, for more complex datasets such as \texttt{lobster}, VR-based methods outperform 2D approaches, highlighting their advantage in selecting thin structures and handling subtle spatial boundaries.

The efficiency metrics reported in Table~\ref{tab:avg-all} further show that Super-Gaussian consistently requires fewer interactions and shorter completion times across all datasets, with especially large improvements in \texttt{vortex} and \texttt{ionization}. Additionally, Table~\ref{tab:avg-pts-per-stroke} reveals that while interaction counts may appear similar in some cases, the average number of points per stroke differs significantly, sometimes by nearly a factor of two, indicating longer sustained interactions (e.g., mid-air gestures). 
\sm{Although this stroke-level measure does not directly quantify fatigue, the reduced number of sampled brush points suggests less sustained mid-air controller motion, which may be relevant to physical effort in prolonged VR selection tasks.}

\sm{Our quantitative selection comparison was conducted by four trained author-operators, which limits its ability to characterize general user behavior. We therefore interpret the results as a controlled technical comparison of selection quality and operational cost, rather than as a usability evaluation.}

\begin{table}[t]
\centering
\caption{Average brush points per stroke (avg pts/st) across the datasets. 
}
\label{tab:avg-pts-per-stroke}
\vspace{-0.1in}
\begin{adjustbox}{width=0.5\columnwidth}
\scriptsize
\setlength{\tabcolsep}{5pt}
\begin{tabular}{lrr}
\toprule
dataset & GSVR & Super-Gaussian \\
\midrule
\texttt{vortex}     & 64.31  & 56.83  \\
\texttt{lobster}    & 48.27  & 39.54  \\
\texttt{ionization} & 364.88 & 156.24 \\
\bottomrule
\end{tabular}
\end{adjustbox}
\vspace{-0.1in}
\end{table}

\begin{table}[t]
\centering
\caption{Rendering efficiency of our editable Gaussian representation and a naive Unity-based DVR baseline in VR. OOM stands for out of memory.}
\label{tab:efficiency}
\vspace{-0.1in}
\begin{adjustbox}{width=\columnwidth}
\scriptsize
\setlength{\tabcolsep}{6pt}
\begin{tabular}{lcccccccccc}
\toprule
dataset 
& \shortstack{volume\\resolution} 
& \shortstack{volume\\size} 
& \shortstack{\# splats\\(MB)} 
& \shortstack{\# base\\scenes} 
& \multicolumn{3}{c}{ours (FPS)} 
& \multicolumn{3}{c}{DVR baseline (FPS)} \\
\cmidrule(lr){6-8}
\cmidrule(lr){9-11}
& & & & 
& avg & min & max 
& avg & min & max \\
\midrule
\texttt{lobster}   
& 602$\times$648$\times$157       
& 234 MB  
& 168,090 (12.0) 
& 1 
& 119.92 & 96.98 & 144.55 
& 119.88 & 100.28 & 139.44 \\

\texttt{aneurysm}  
& 512$\times$512$\times$512       
& 512 MB  
& 132,490 (9.6)  
& 1 
& 120.01 & 97.05 & 145.35 
& 75.18 & 3.95 & 131.70 \\

\texttt{bonsai}    
& 512$\times$512$\times$512       
& 512 MB  
& 831,612 (57.9) 
& 2 
& 119.88 & 93.02 & 138.42 
& 66.92 & 0.84 & 170.66 \\

\texttt{beetle}    
& 864$\times$864$\times$494       
& 1.37 GB 
& 266,808 (18.7) 
& 1 
& 120.01 & 98.88 & 135.11 
& \multicolumn{3}{c}{OOM} \\

\texttt{vortex}    
& 1,000$\times$1,000$\times$1,000 
& 3.73 GB 
& 164,721 (11.8) 
& 2 
& 118.10 & 62.48 & 148.86 
& \multicolumn{3}{c}{OOM} \\

\texttt{Nyx}       
& 1,024$\times$1,024$\times$1,024 
& 4 GB    
& 169,737 (12.6) 
& 2 
& 118.00 & 61.42 & 189.00 
& \multicolumn{3}{c}{OOM} \\

\texttt{chameleon} 
& 1,024$\times$1,024$\times$1,080 
& 4.22 GB 
& 372,702 (25.8) 
& 2 
& 119.93 & 99.54 & 130.14 
& \multicolumn{3}{c}{OOM} \\
\bottomrule
\end{tabular}
\end{adjustbox}
\vspace{-0.1in}
\end{table}

\vspace{-0.075in}
\subsection{Rendering Efficiency}

To evaluate the rendering efficiency of our editable Gaussian representation in VR, Table~\ref{tab:efficiency} reports the memory footprint of each dataset relative to its original volume size, along with the corresponding rendering performance. FPS is measured by traversing viewpoints along a 360° camera pan, following a full warm-up rotation to stabilize the rendering pipeline. Across all datasets, our method consistently achieves an average FPS close to 120, well above the 90 FPS threshold for VR comfort~\cite{laviola2000discussion, scholl2018direct}, regardless of the number of splats or volume resolution. This demonstrates the strong compression capability of Gaussian-splatting-based VolVis, in which even multi-gigabyte volumes, such as the 4 GB \texttt{Nyx} dataset, are rendered at real-time frame rates using only 12.6 MB of splats. However, the minimum and maximum FPS vary across datasets depending on the number of splats and scene complexity, indicating that rendering cost fluctuates under specific camera angles and viewpoint configurations. \sm{The practical scalability of the framework is primarily determined by the number of Gaussian primitives rather than the original volume resolution.}

\sm{We further compare our editable Gaussian representation with a naive DVR baseline under the same VR benchmarking protocol. For the baseline, we used the open-source UnityVolumeRendering's DVR mode with Phong lighting~\cite{lavik_unityvolumerendering_2024}.
The sampling rate multiplier was set to 1.0, and early ray termination was enabled, terminating ray marching once the accumulated opacity exceeded 0.996. In this baseline, DVR performance degrades noticeably as the volume resolution increases. While DVR remains feasible for smaller datasets, its frame rate drops even for $512^3$ volumes, and datasets larger than \texttt{beetle} could not be rendered due to Unity's 3D texture allocation limit. In particular, for the $512^3$ datasets (\texttt{aneurysm}, \texttt{bonsai}), the minimum DVR frame rate falls below $5$ FPS in some views, indicating severe view-dependent performance fluctuation. Even when the average FPS exceeds $60$, such extreme drops would likely cause discomfort in VR and degrade the interactive user experience. 
Overall, these measurements show that the editable Gaussian representation provides a compact and stable rendering substrate for our interactive VR workflow. The DVR baseline highlights the practical challenges of applying a standard Unity-based ray-marching renderer to large volumetric datasets in the same environment. However, this comparison should be interpreted within the scope of this baseline implementation, which does not include advanced acceleration strategies such as parallel rendering, out-of-core data management, foveated rendering, or adaptive sampling/downsampling. A comprehensive comparison with highly optimized DVR pipelines remains outside the scope of this study.}

\begin{figure}[t]
\centering
\includegraphics[width=\linewidth]{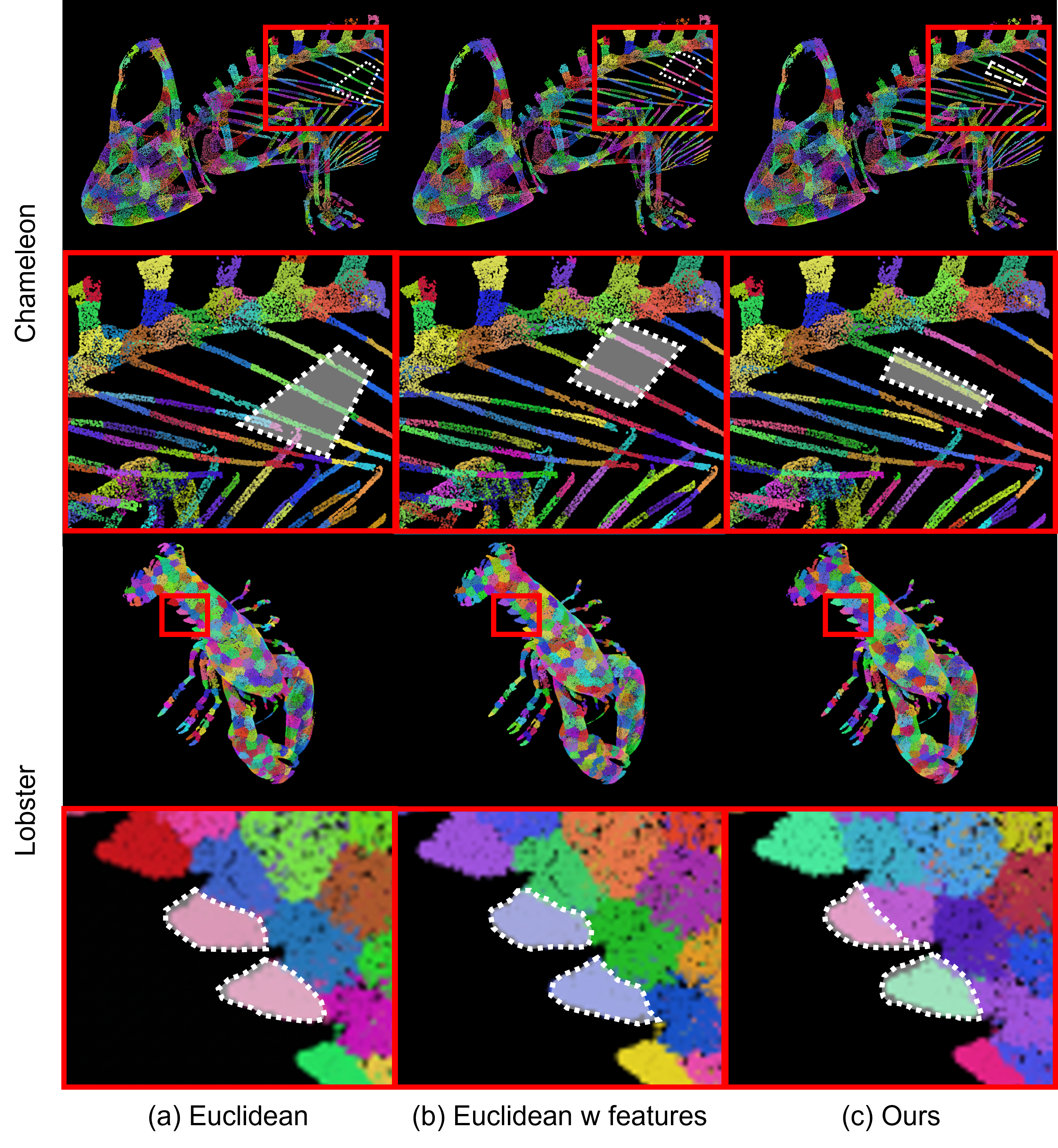}
\vspace{-0.25in}
\caption{
Ablation on distance formulation in SLIC-based Super-Gaussian construction.
The top row shows the \texttt{chameleon} dataset, and the bottom row shows the \texttt{lobster} dataset.
Clusters are visualized with random colors and rendered with Gaussian centroids.
(a) Euclidean distance, (b) Euclidean distance with feature similarity, and (c) ours (geodesic distance with feature similarity). 
}
\label{fig:slic_ablation_dist}
\vspace{-0.1in}
\end{figure}

\vspace{-0.05in}
\subsection{\sm{Geodesic SLIC Ablation}}
\label{app:slic}
\sm{We conduct an ablation study on Super-Gaussian construction to evaluate the effect of distance formulation in our SLIC-based clustering. All experiments are performed under identical settings, including the same number of clusters and shared hyperparameters such as voxel-based seed initialization, compactness, and graph construction, applied to the same optimized Gaussian sets for each dataset. As Super-Gaussian is designed for interaction-oriented over-segmentation rather than final semantic segmentation, we evaluate clustering quality qualitatively.
We fix the feature representation to our proposed geometric feature set and vary the distance formulation. We compare geodesic distance with feature similarity against Euclidean distance with and without feature contribution, where the latter corresponds to a purely spatial SLIC-style baseline. As shown in Fig.~\ref{fig:slic_ablation_dist}, Euclidean distance-based clustering tends to group spatially adjacent Gaussians regardless of structural continuity. For example, in the \texttt{chameleon} dataset, clusters fail to follow the rib structures and instead merge neighboring regions across ribs. In contrast, our method preserves the directional continuity of the ribs. A similar pattern is observed in the \texttt{lobster} dataset, where Euclidean distance merges adjacent regions near the curved shell tip, while our method successfully separates clusters along the shell structure.}




\vspace{-0.075in}
\section{Limitations and Future Work}

The quality of Super-Gaussian depends on preprocessing and hyperparameter choices. Our method relies on precomputed clusters, where parameters such as the SLIC window size $S$ directly control the clustering granularity. While we adopt an over-segmentation strategy to allow flexible user-driven selection, the resulting cluster size and boundary quality can vary significantly across datasets. In particular, regions with weak structural contrast or ambiguous boundaries may lead to inconsistent clustering results. Furthermore, the effectiveness of RW-based selection depends on the quality and distribution of user-provided seeds. Sparse or ambiguous seed placement may lead to inaccurate propagation, particularly in regions with complex topology or weak feature separation. Although our select-and-refine workflow mitigates this issue through subsequent corrections, the initial RW result can still affect the overall efficiency of interactions. A promising direction is to incorporate learning-based approaches to guide seed placement or propagation, enabling more robust and user-independent selection behavior. We will explore adaptive or data-driven clustering that automatically adjusts cluster granularity based on local structural properties, and a learning-based solution to guide seed placement or propagation, thereby achieving more robust, user-independent selection behavior.

\sm{Our framework also depends on the quality of the initial multi-view renderings or TF-specific scenes used to construct the Gaussian representation. When such scenes are not carefully curated to isolate specific structures, our select-and-refine workflow allows users to subsequently separate and edit regions through interaction, mitigating the impact of suboptimal initial scenes to some extent. However, the resulting editable Gaussian set still 
inherits limitations from the initial scene and TF design, and our framework does not eliminate this dependency. A promising direction for future work is to automate the construction of multiple TF-specific base representations by combining volume rendering with image-space segmentation methods~\cite{ai2025nli4volvis}, reducing reliance on manually curated input scenes.
}

\sm{Another limitation arises from the primitive-level nature of our selection. Since Super-Gaussian performs selection at the Gaussian-primitive level, a single Gaussian may occasionally cover the boundary between multiple overlapping or adjacent structures. In such cases, the Gaussian is assigned to only one cluster or editable unit, and the resulting selection boundary may not perfectly match the actual object boundary. This can introduce gaps or visual artifacts when selected splats are removed, isolated, or filtered, especially when large Gaussians cover multiple semantic regions. This boundary ambiguity is an inherent limitation of Gaussian-level segmentation and selection, and is also a common issue in Gaussian-splatting-based segmentation methods. As a future direction, a post-refinement strategy after selection or segmentation would be needed to reduce such artifacts. Since our target regions are user-defined and interactively determined, boundary refinement after selection is preferable to assuming predefined semantic boundaries during training. For example, adaptively reducing the size of boundary Gaussians or splitting ambiguous Gaussians into smaller primitives could help reduce such artifacts.}

Finally, our semantic representation relies on CLIP embeddings without additional fine-tuning. The current approach combines image and text embeddings through simple averaging to handle newly added labels in real time. However, CLIP is primarily trained on natural images, and its performance may degrade when applied to volume-rendered images with domain-specific characteristics. While fine-tuning CLIP on volumetric data could improve semantic alignment, it would introduce additional computational overhead and conflict with our goal of real-time labeling. Future work includes exploring lightweight adaptation strategies or domain-specific embedding models that better capture the semantics of volumetric data while preserving real-time interaction.

\vspace{-0.075in}
\section{Conclusions}

We have presented Super-Gaussian, a structure-aware interaction framework for Gaussian-splatting-based VolVis in immersive VR. By grouping Gaussian primitives into feature-aware units and introducing a hierarchical select-and-refine workflow, our method enables efficient and flexible ROI specification without relying on pre-segmentation.
Our framework integrates spatial interaction, Gaussian-based rendering, and natural language interfaces within a unified VPA loop, allowing users to iteratively explore, select, and semantically interpret volumetric structures. Through case studies and user evaluations, we demonstrated that our method improves selection accuracy, reduces interaction effort, and enables more effective exploration than existing approaches.

%
%
%
%
\vspace{-0.1in}
\acknowledgments{
This work was supported in part by the National Research Foundation of Korea under Grant RS-2024-00349697 and Grant RS-2021-NR060143; in part by the Institute for Information and Communications Technology Planning and Evaluation under Grant IITP-2026-RS-2020-II201819; in part by the National Research Council of Science \& Technology(NST) grant by the Korea government (MSIT) (No. GTL24033-000); and in part by Korea University Grant.
K.\ Tang and C.\ Wang are supported by funding from the U.S.\ National Science Foundation IIS-2101696, OAC-2104158, IIS-2401144, and CCF-2550610, and the U.S.\ Department of Energy DE-SC0023145.
The authors thank the anonymous reviewers for their insightful comments.
}
\bibliographystyle{abbrv-doi-hyperref}
\vspace{-0.1in}
\bibliography{template}

\appendix 
\crefalias{section}{appendix} 

\newpage
\clearpage

\setcounter{section}{0}
\setcounter{figure}{0}
\setcounter{table}{0}

\section{User Interface of Super-Gaussian}

\begin{figure}[ht]
\centering
\includegraphics[width=0.8\linewidth]{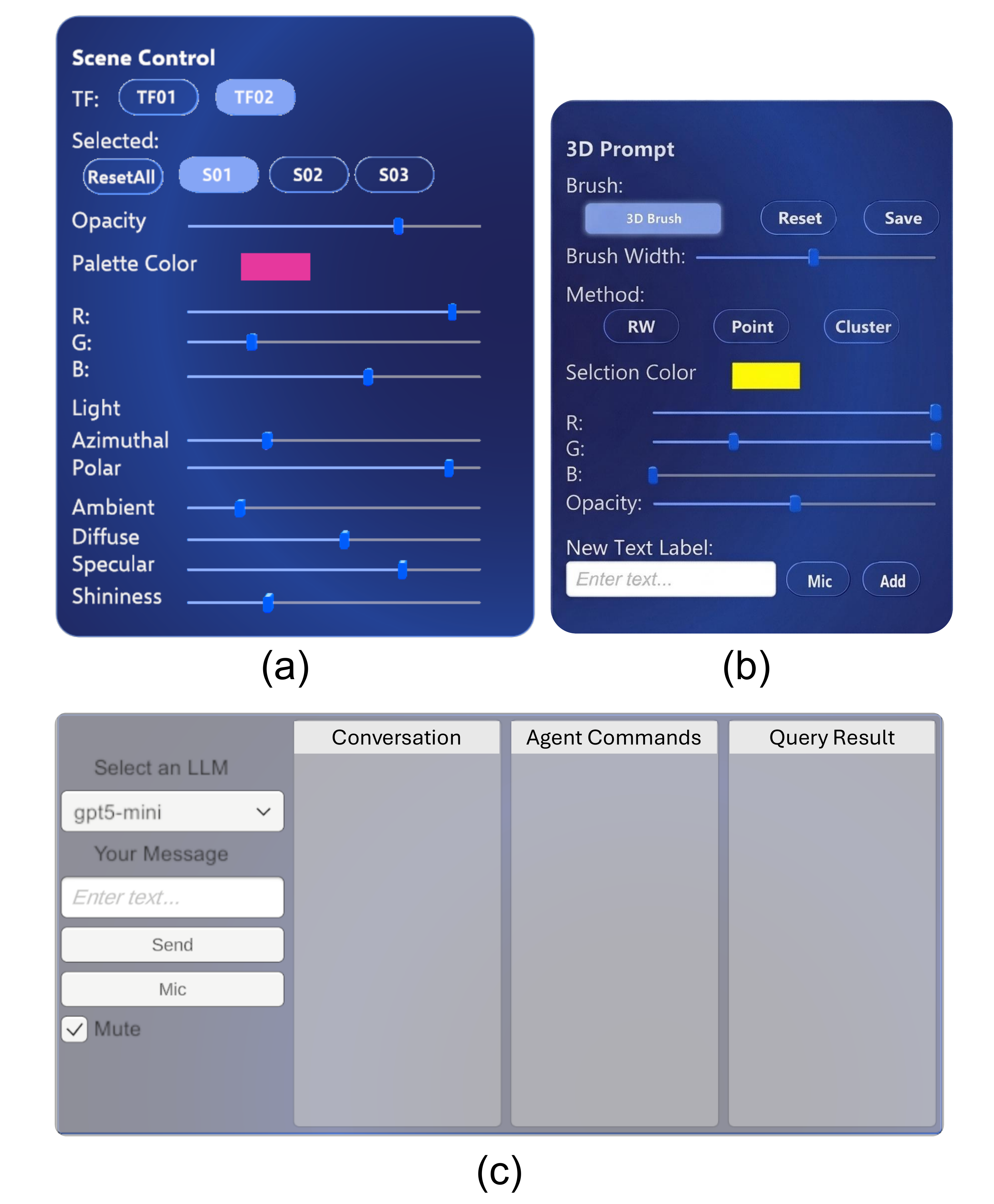}
\vspace{-0.1in}
\caption{User interface of Super-Gaussian in the VR environment. (a) Scene control panel for GUI-based Scene editing. (b) 3D prompt panel for brush interaction, selection control, and semantic labeling. (c) NLI interface for voice-command-based volume exploration and scene editing.
}
\label{fig:appendix-ui}
\end{figure}

Super-Gaussian provides three main user interfaces. Two of them are designed for direct GUI-based scene control and editing, while the third supports NLI, allowing users to follow the conversation and inspect intermediate results, such as intent parsing and open-vocabulary queries.
Specifically, the scene control panel in Fig.~\ref{fig:appendix-ui} (a) allows users to edit visualization attributes of editable units, including color, opacity, and lighting parameters.
The 3D prompt panel in Fig.~\ref{fig:appendix-ui} (b) supports interactive region selection and labeling. Users can adjust brush width, choose selection methods (RW, cluster, or point), and modify the color and opacity for highlighting selected regions. It also provides GUI-based text labeling and options to either save the selection as a new editable unit in Fig.~\ref{fig:appendix-ui} (a) or reset the current selection.
The NLI interface in Fig.~\ref{fig:appendix-ui} (c) provides a conversational interface where users can review voice commands and system responses in text form. It also includes an Agent Commands panel to inspect intermediate agent operations, such as intent parsing and execution steps, and a Query Result panel to display the results of open-vocabulary queries.

\vspace{-0.05in}
\section{Editable Gaussian Features}
\label{app:features}
\begin{figure}[t]
\centering
\includegraphics[width=\linewidth]{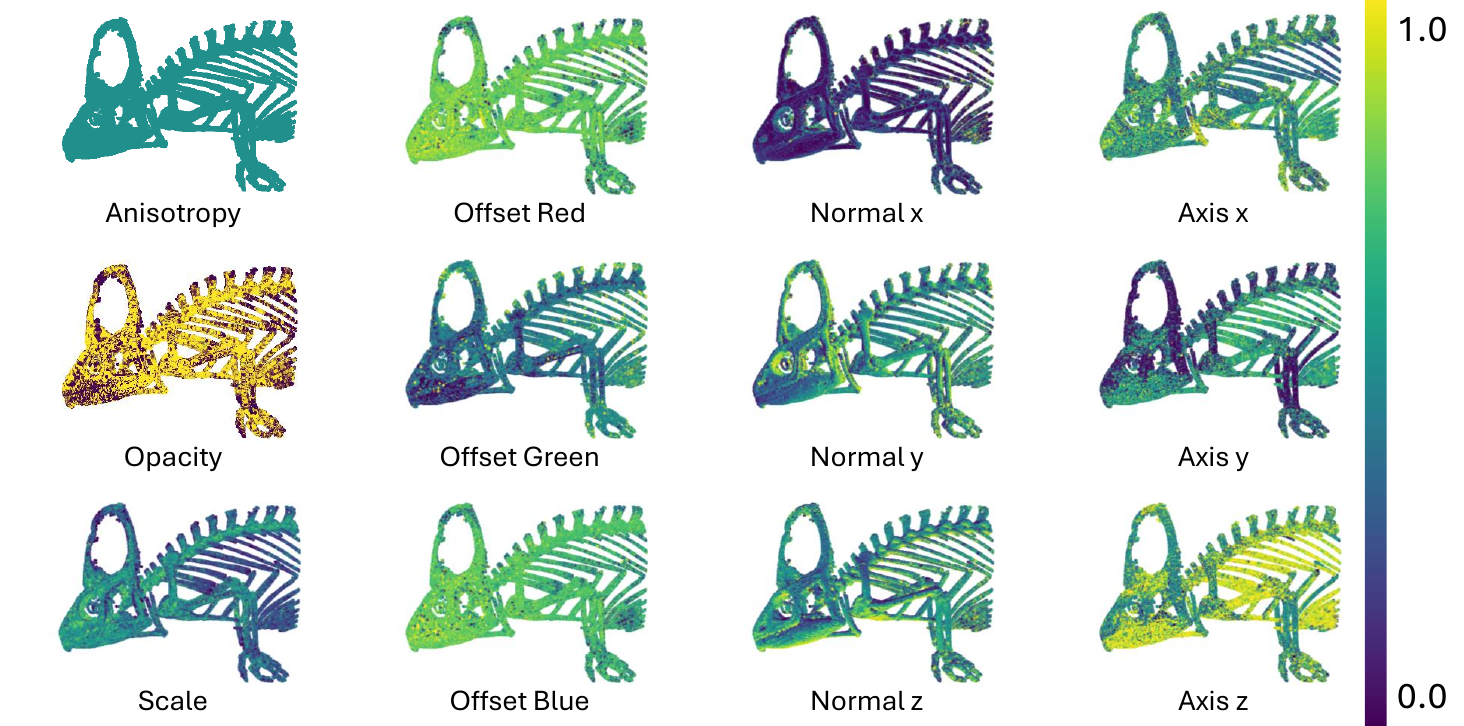}
\vspace{-0.25in}
\caption{Gaussian attributes of the \texttt{chameleon} dataset color-mapped onto the 3D scene to illustrate its spatial distribution, including anisotropy, opacity, scale, offset color, normal directions, and principal axes.
}
\label{fig:feature}
\end{figure}
Editable Gaussian primitives are parameterized by both geometric and appearance-related attributes.
Geometric attributes include anisotropy, scale, normal directions, and principal axes, which describe the local structure and orientation of each Gaussian.
Appearance-related attributes include opacity and offset color, which control the volume's visual representation.
As illustrated in Fig.~\ref{fig:feature}, these attributes exhibit distinct spatial distributions across the volume.
In particular, geometric attributes tend to form coherent patterns aligned with underlying structures, while appearance-related attributes often show less structural consistency.
These properties are important for our method, as structure-aware attributes (e.g., scale, normal, and principal axes) guide Super-Gaussian clustering and enable more reliable region selection.

\vspace{-0.05in}
\section{Random Walk Ablation}
\label{app:rw}
\begin{figure}[ht]
\centering
\includegraphics[width=1.0\linewidth]{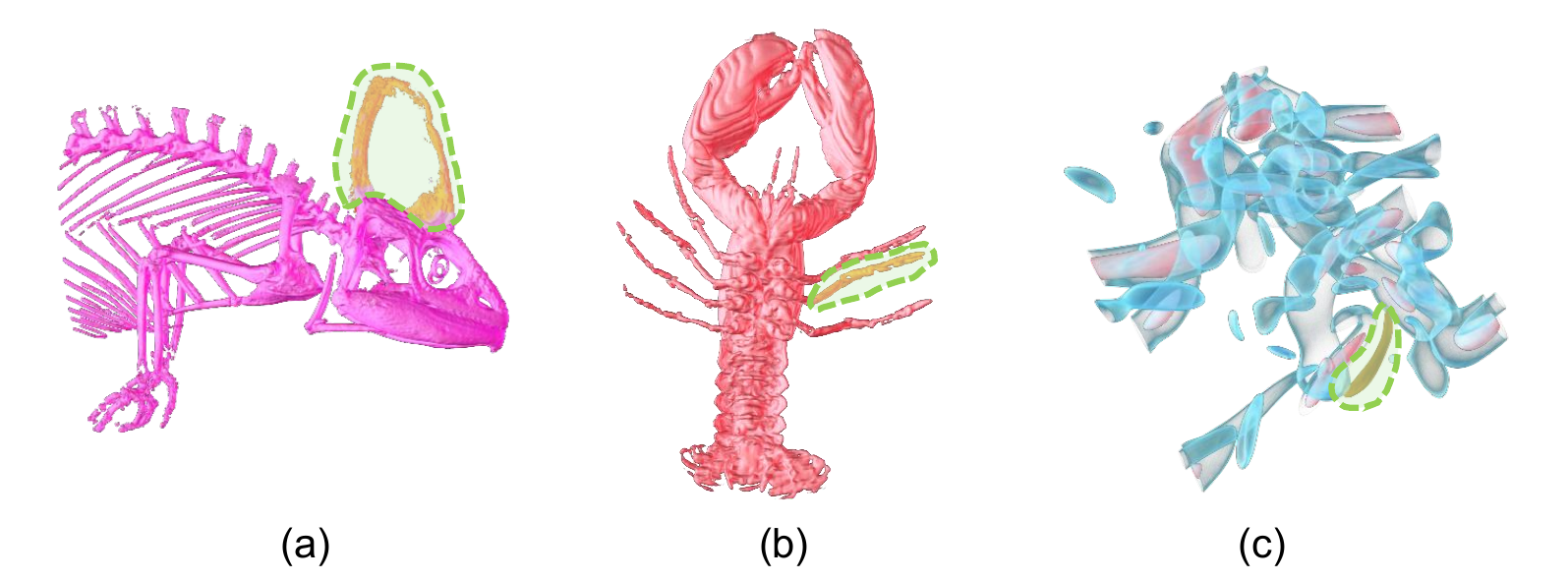}
\vspace{-0.25in}
\caption{Datasets used in the RW ablation: (a) \texttt{chameleon}, (b) \texttt{lobster}, and (c) \texttt{vortex}. The target selected is highlighted with a green dotted line.}
\label{fig:rw-ablation}
\end{figure}

We conduct an ablation study on our RW selection mode using the \texttt{chameleon}, \texttt{lobster}, and \texttt{vortex} datasets. Note that RW is sensitive to the choice of foreground and background seeds. Therefore, the results should be interpreted as a reference rather than absolute performance. 
Table~\ref{tab:rw_comparison} compares our method with Gaussian RW (i.e., RW performed on individual Gaussians) and cluster RW using Euclidean distance under identical brush inputs. The reported metrics (IoU, accuracy, F1 score, and precision) are measured for the target region shown in Fig.~\ref{fig:rw-ablation}, and execution time is reported in seconds.
Gaussian RW achieves strong segmentation accuracy in some cases (e.g., \texttt{vortex}), but shows inconsistent performance across datasets and incurs significantly higher computational cost, requiring up to two orders of magnitude longer execution time than our method. In contrast, cluster RW using Euclidean distance provides faster execution, but leads to noticeably lower accuracy, particularly in IoU and F1 score. This highlights the importance of structure-aware distance, as Euclidean distance fails to capture underlying spatial connectivity in volumetric data. Our method consistently achieves high segmentation quality across all datasets while maintaining efficient runtime, demonstrating a better balance between accuracy and efficiency.

\begin{table}[t]
\centering
\caption{Comparison of RW variants across datasets.}
\label{tab:rw_comparison}
\vspace{-0.1in}
\begin{adjustbox}{width=\columnwidth}
\begin{tabular}{llccccc}
\toprule
dataset & method & time (s) & IoU & acc. & F1 & prec. \\
\midrule
\multirow{3}{*}{\texttt{chameleon}}
& ours & 0.0807 & 0.8814 & 0.9977 & 0.8814 & 1.0000 \\
& cluster RW (Euclidean) & 0.0217 & 0.5538 & 0.9854 & 0.5538 & 1.0000 \\
& Gaussian RW (geodesic) & 9.9769 & 0.5327 & 0.9850 & 0.5327 & 1.0000 \\
\midrule
\multirow{3}{*}{\texttt{lobster}}
& ours & 0.0833 & 0.9012 & 0.9980 & 0.9018 & 0.9993 \\
& cluster RW (Euclidean) & 0.0205 & 0.5538 & 0.9854 & 0.5538 & 1.0000 \\
& Gaussian RW (geodesic) & 11.5861 & 0.8622 & 0.9971 & 0.8622 & 1.0000 \\
\midrule
\multirow{3}{*}{\texttt{vortex}}
& ours & 0.6044 & 0.5247 & 0.9850 & 0.6883 & 1.0000 \\
& cluster RW (Euclidean) & 0.0903 & 0.1036 & 0.7273 & 0.1877 & 0.1036 \\
& Gaussian RW (geodesic) & 4.7902 & 1.0000 & 1.0000 & 1.0000 & 1.0000 \\
\bottomrule
\end{tabular}
\end{adjustbox}
\end{table}

\vspace{-0.05in}
\section{\sm{Clustering Robustness}}




\begin{table*}[t]
\centering
\caption{\sm{Comparison of Gaussian-based scene representations used in our evaluation.}}
\label{tab:gs_methods}
\vspace{-0.1in}
\begin{adjustbox}{width=\linewidth}
\begin{tabular}{lllll}
\toprule
method & primitive & initialization & optimization & distinctive feature \\
\midrule

3DGS
& explicit 3D ellipsoids
& sparse point cloud
& clone / split / prune
& SH-based view-dependent appearance
\\

2DGS
& explicit 2D surfels
& sparse point cloud
& surfel densification + depth/normal regularization
& surface-consistent geometry
\\

Scaffold-GS
& voxel anchors + neural Gaussians
& sparse voxel anchors
& anchor growing / pruning
& MLP-generated Gaussians
\\

iVR-GS
& Gaussians + material attributes
& sparse point cloud
& joint geometry / material optimization
& editable transfer function \& relighting
\\

\bottomrule
\end{tabular}
\end{adjustbox}
\end{table*}

\begin{figure}[ht]
\centering
\includegraphics[width=0.9\linewidth]{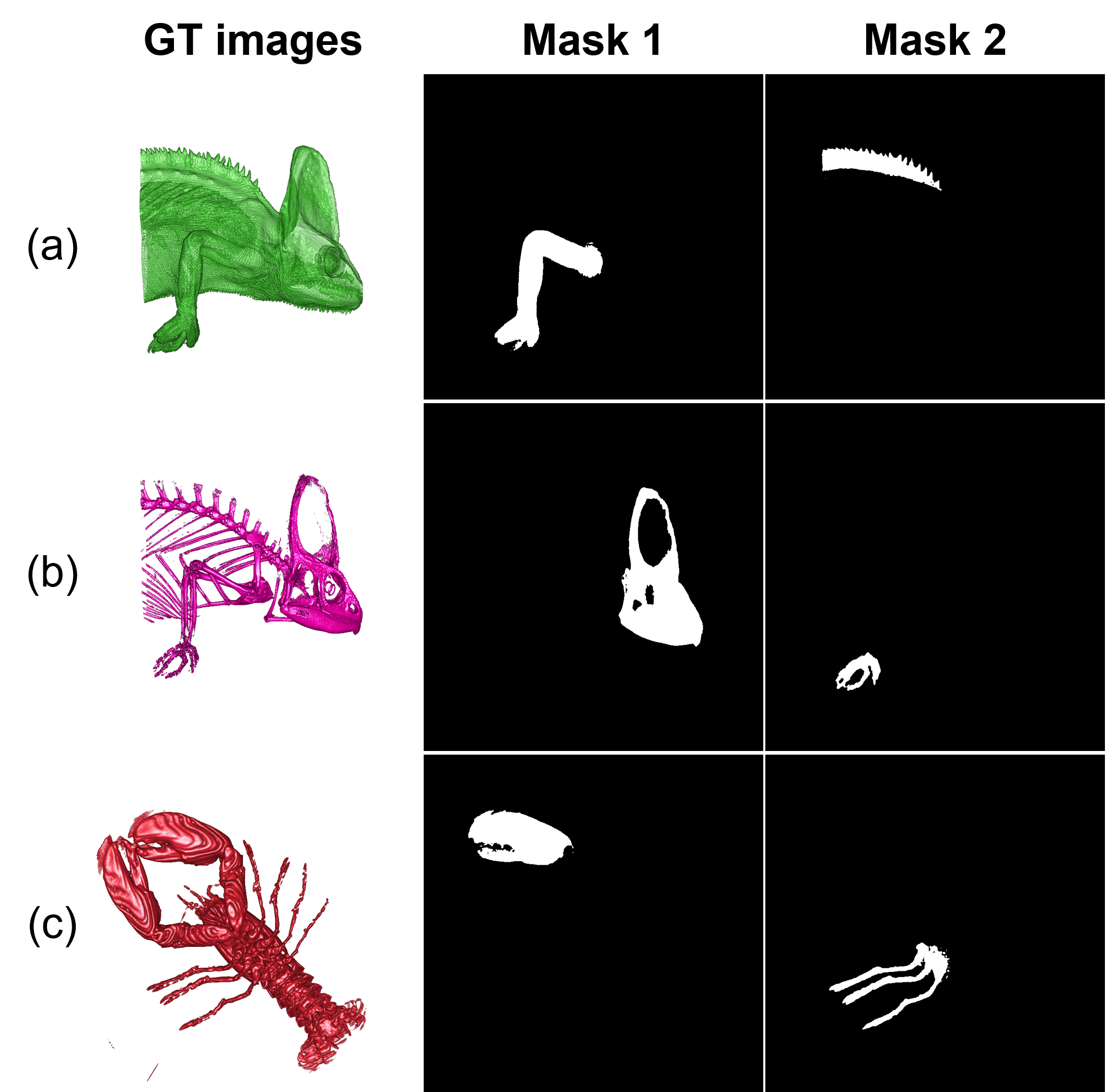}
\vspace{-0.1in}
\caption{Datasets used in the clustering robustness experiments: (a) \texttt{chameleon skin}, (b) \texttt{chameleon bone}, and (c) \texttt{lobster}.}
\label{fig:robust_mask}
\end{figure}

\begin{table}[t]
\centering
\caption{\sm{Comparison of Super-Gaussian clustering quality across different Gaussian scene representations. Lower is better for mean size, size coefficient of variation (CV), and compactness, while higher is better for same-label kNN.}}
\label{tab:cluster_generalization}
\vspace{-0.1in}
\resizebox{\linewidth}{!}{
\begin{tabular}{llccccc}
\toprule
dataset & method & \# clusters & mean size & size CV & compactness & same-label kNN \\
\midrule

\multirow{4}{*}{\texttt{chameleon skin}}
& 2DGS        & 495 & 4143.08 & 1.8833 & 0.0128 & 0.9272 \\
& 3DGS        & 458 & 637.66  & 0.6483 & 0.0169 & 0.9293 \\
& iVR-GS      & 472 & 574.60  & 0.6825 & 0.0173 & 0.9131 \\
& Scaffold-GS & 191 & 152.71  & 0.4854 & 0.0312 & 0.8972 \\

\midrule

\multirow{4}{*}{\texttt{chameleon bone}}
& 2DGS        & 367 & 281.57 & 0.4480 & 0.0167 & 0.9530 \\
& 3DGS        & 374 & 802.88 & 0.5837 & 0.0164 & 0.9694 \\
& iVR-GS      & 388 & 702.47 & 0.6925 & 0.0159 & 0.9669 \\
& Scaffold-GS & 182 & 136.17 & 0.3789 & 0.0267 & 0.9356 \\

\midrule

\multirow{4}{*}{\texttt{lobster}}
& 2DGS        & 393 & 216.27 & 0.3751 & 0.0137 & 0.9117 \\
& 3DGS        & 412 & 488.32 & 0.5480 & 0.0124 & 0.9417 \\
& iVR-GS      & 418 & 465.26 & 0.6049 & 0.0122 & 0.9341 \\
& Scaffold-GS & 118 & 149.23 & 0.3504 & 0.0283 & 0.9202 \\

\bottomrule
\end{tabular}
}
\end{table}

\sm{To evaluate the robustness of our clustering algorithm across different Gaussian scene representations, we trained four representative Gaussian-based methods (3DGS~\cite{kerbl2023gs}, 2DGS~\cite{huang20242d}, Scaffold-GS~\cite{lu2024scaffold}, and iVR-GS~\cite{tang2025ivrgs} (ours), see Table~\ref{tab:gs_methods}) on the same scenes. 
We generated Super-Gaussians using the proposed clustering algorithm. 
Datasets without semantic boundaries (e.g., simulation datasets) make it difficult to evaluate clustering robustness. 
Therefore, we conducted experiments on the three scenes shown in Fig.~\ref{fig:robust_mask}: \texttt{chameleon skin}, \texttt{chameleon bone}, and \texttt{lobster}.

Since Super-Gaussians are intentionally designed as over-segmented interaction units, no semantic or object-level ground truth exists for direct evaluation. 
We therefore first analyze the intrinsic clustering quality independent of semantic masks.
For a fair comparison, all methods use the same geometric feature representation described in Section~\ref{sec:supergs}, consisting of the Gaussian mean scale, normal direction, and local principal axes. 
Since Scaffold-GS generates view-dependent Gaussians through neural decoding, clustering is instead performed on the anchor primitives using the corresponding anchor features.

Table~\ref{tab:cluster_generalization} summarizes several intrinsic clustering metrics. 
Specifically, mean size measures the average number of Gaussians per cluster (smaller indicates finer over-segmentation), size coefficient of variation (CV) denotes the coefficient of variation of cluster sizes, which measures cluster size regularity (lower is better), compactness evaluates the normalized spatial radius of each cluster (lower is better), and same-label kNN measures local neighborhood consistency (higher is better).
%
Overall, the proposed clustering algorithm produces stable clustering quality with different trade-offs across all Gaussian representations despite their substantially different primitive definitions. 
While Scaffold-GS naturally produces fewer clusters due to its sparse anchor-based representation, its clusters remain spatially coherent with competitive size regularity. 
Similarly, both 3DGS and iVR-GS exhibit comparable compactness and local consistency, despite iVR-GS having additional editable attributes. 
Although 2DGS occasionally produces larger cluster sizes owing to its surface-oriented primitive distribution, it maintains competitive spatial compactness and neighborhood coherence. 
These results indicate that the proposed feature representation and graph-based clustering are largely independent of the underlying Gaussian representation and generalize well across different editable Gaussian scene models.

To further evaluate whether these over-segmented Super-Gaussians can effectively reconstruct semantic regions, we additionally generate pseudo-semantic masks (Fig.~\ref{fig:robust_mask} Mask1, 2) using SAM2~\cite{ravi2025sam} and measure the boundary agreement between the union of selected Super-Gaussians and pseudo masks. 
The reconstruction quality is evaluated by measuring the intersection-over-union (IoU) between the reconstructed region and the corresponding pseudo-semantic mask. 
Formally, let $M_{\mathrm{SAM}}$ denote the binary pseudo mask generated by SAM2 and  $M_{\mathrm{SG}}=\bigcup_{i\in\mathcal{S}} G_i$ the union of the selected Super-Gaussians, where $\mathcal{S}$ is the set of selected clusters.}
\sm{
The IoU is computed as
\[
\mathrm{IoU}(M_{\mathrm{SAM}}, M_{\mathrm{SG}})
=
\frac{|M_{\mathrm{SAM}} \cap M_{\mathrm{SG}}|}
{|M_{\mathrm{SAM}} \cup M_{\mathrm{SG}}|}.
\]
}
\sm{
The quantitative results are given in Table~\ref{tab:iou}. Across 2DGS, 3DGS, and iVR-GS, IoU values remain consistently around 0.85, indicating that Super-Gaussian clustering robustly reconstructs semantic regions regardless of the underlying Gaussian representation. 
Although small variations exist across datasets, these differences are minor and unlikely to reflect meaningful differences in segmentation quality, particularly because the evaluation is based on pseudo-masks rather than manually annotated ground truth. 
In contrast, ScaffoldGS consistently exhibits noticeably lower IoU. This behavior is expected because ScaffoldGS first groups Gaussians into anchor-based structures, and Super-Gaussian clustering is consequently performed at the anchor level rather than directly on individual Gaussian primitives. The reduced spatial granularity limits the precision of reconstructed semantic boundaries, leading to lower boundary agreement with the pseudo masks.}

\begin{table}[htb]
\centering
\caption{\sm{IoU comparison of semantic region reconstruction. IoU is measured between the union of the selected super Gaussians and the corresponding SAM2 pseudo mask (higher is better).}}
\label{tab:iou}
\vspace{-0.1in}
\resizebox{0.8\linewidth}{!}{
\begin{tabular}{lcccc}
\toprule
dataset & 2DGS & 3DGS & iVR-GS & ScaffoldGS \\
\midrule
\texttt{chameleon skin} & {0.8620} & 0.8606 & 0.8531 & 0.7321 \\
\texttt{chameleon bone} & 0.8517 & {0.8546} & 0.8471 & 0.8417 \\
\texttt{lobster}   & 0.8392 & 0.8577 & {0.8596} & 0.7966 \\
\bottomrule
\end{tabular}
}
\end{table}

\vspace{-0.05in}
\section{Text Label Latency}
\label{app:text}
We report the latency of real-time text labeling in Table~\ref{tab:text-latency}. While user-perceived latency may vary with system conditions, we observe a consistent trend: latency increases slightly as the number of selected points increases. This is expected, as the selected Gaussians are rasterized into 2D before extracting CLIP image embeddings, and the cost of this 2D rasterization step scales with the number of selected Gaussians. Consequently, the rendering overhead introduced during this preprocessing stage appears to be the primary factor affecting latency. The 2D rasterization and CLIP embedding extraction are performed on a local Python server. All components run on a single local machine, and communication between the PC-VR client and the Python server is handled via a RESTful API over the local loopback interface.

\begin{table}[htb]
\centering
\caption{Text labeling latency (in seconds) with respect to the number of selected Gaussians ($10^2$, $10^3$, and  $10^4$).}
\label{tab:text-latency}
\vspace{-0.1in}
\begin{adjustbox}{width=0.4\columnwidth}
\scriptsize
\setlength{\tabcolsep}{4pt}
\begin{tabular}{lrrr}
\toprule
& \multicolumn{3}{c}{Selected Gaussians} \\
\cmidrule(lr){2-4}
Dataset & $10^2$ & $10^3$ & $10^4$ \\
\midrule
\texttt{chameleon} & 12.5 & 13.8 & 15.0 \\
\texttt{beetle}    & 11.9 & 13.7 & 14.5 \\
\texttt{lobster}   & 11.4 & 12.8 & 13.8 \\
\texttt{Nyx}       & 11.2 & 12.1 & 13.9 \\
\texttt{aneurysm}  & 11.2 & 11.8 & 12.7 \\
\bottomrule
\end{tabular}
\end{adjustbox}
\end{table}

\vspace{-0.05in}
\section{\sm{Super-Gaussian for Unstructured Data}}

\begin{figure}[ht]
\centering
\includegraphics[width=1.0\linewidth]{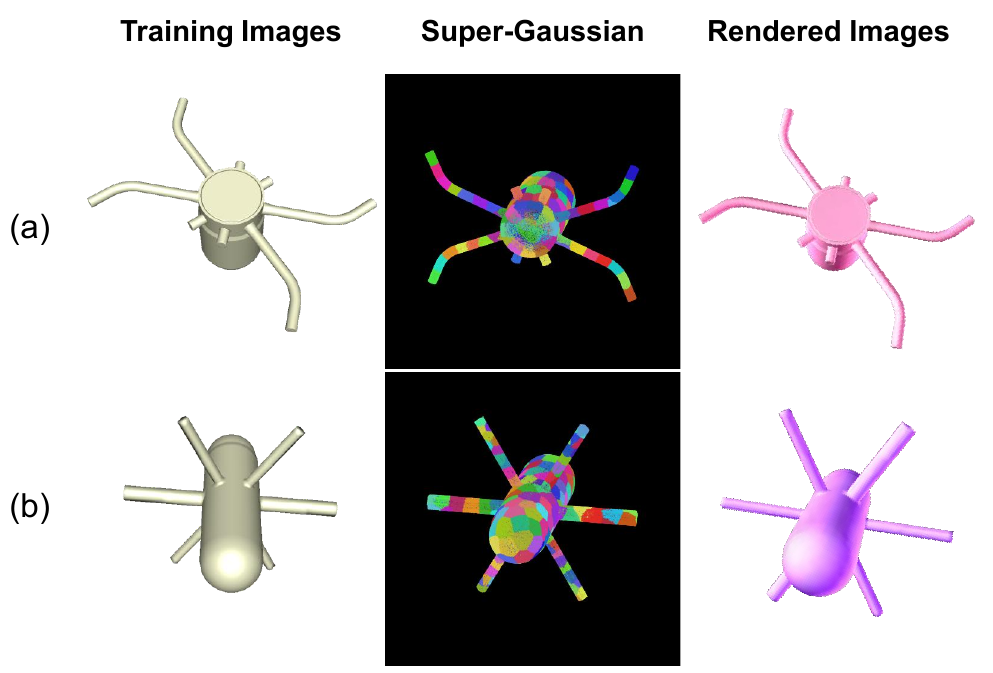}
\vspace{-0.25in}
\caption{\sm{Unstructured dataset results. From left to right: training images for editable 3D Gaussian Splatting, Super-Gaussian clustering, and VR rendering results. (a) \texttt{ROCOM} and (b) \texttt{CUPID}.}}
\label{fig:unstructured}
\end{figure}

\sm{Although our work primarily targets volume visualization, the proposed framework is not fundamentally restricted to structured volumetric grids. Since Super-Gaussian clustering and ROI interaction operate directly on editable Gaussian representations rather than the original volume, the same pipeline can be applied to unstructured datasets once they are converted into editable Gaussian models through multi-view rendering and Gaussian optimization.

To demonstrate this applicability, we trained editable Gaussian models on two representative unstructured scientific datasets, \texttt{ROCOM} and \texttt{CUPID}, and applied our Super-Gaussian clustering without any modification~\cite{cho2019numerical}.
\texttt{ROCOM} is an experimental multivariate reactor coolant mixing dataset defined on an unstructured tetrahedral mesh, containing 12.4 million spatial samples over 5 timesteps with 8 physical variables.
\texttt{CUPID} is a multivariate time-varying thermal-hydraulic simulation dataset of a nuclear reactor, represented on an unstructured tetrahedral mesh, containing 11.3 million spatial samples over 30 timesteps with 11 physical variables.

Fig.~\ref{fig:unstructured} presents the resulting editable Gaussian representations together with the corresponding Super-Gaussian clustering results. Despite being trained on unstructured tetrahedral meshes rather than structured volumes, our method successfully produces spatially coherent super Gaussians that can be used directly as editable interaction units. These results suggest that the proposed framework generalizes beyond structured volumetric datasets and can naturally support unstructured scientific data represented as editable Gaussian models.}

\end{document}